\documentclass[12pt,compress]{elsarticle}
\usepackage{amsmath,amssymb,amscd}
\usepackage{color}
\journal{Nuclear Physics B}

\newcommand{\beq}{\begin{equation}}
\newcommand{\eeq}{\end{equation}}
\newcommand{\bea}{\begin{eqnarray}}
\newcommand{\eea}{\end{eqnarray}}
\newcommand{\bse}{\begin{subequations}}
\newcommand{\ese}{\end{subequations}}

\newcommand{\noi}{\noindent}

\newcommand{\nn}{\nonumber}

\newcommand{\ba}{\begin{array}}
\newcommand{\ea}{\end{array}}
\newcommand{\balign}{\begin{align}}
\newcommand{\ealign}{\end{align}}

\newcommand{\al}{\alpha}
\newcommand{\be}{\beta}

\newcommand{\ep}{\epsilon}
\newcommand{\vep}{\varepsilon}

\newcommand{\si}{\sigma}

\newcommand{\La}{\Lambda}

\newcommand{\mbf}[1]{\mathbf{#1}}

\newcommand{\mbs}[1]{\boldsymbol{#1}}

\newcommand{\mbb}[1]{\mathbb{#1}}

\newcommand{\mc}[1]{\mathcal{#1}}

\newcommand{\mr}[1]{\mathrm{#1}}

\newcommand{\wt}[1]{\widetilde{#1}}

\newcommand{\wh}[1]{\widehat{#1}}

\newcommand{\one}{1\hskip -1mm{\rm l}}

\renewcommand{\d}{\partial}

\renewcommand{\leq}{\leqslant}
\renewcommand{\geq}{\geqslant}

\renewcommand{\r}{\rangle}
\renewcommand{\v}{\vert}
\newcommand{\half}{$\frac{1}{2}~$}
\newcommand{\su}[1]{su($#1$)}

\newcommand\ket[1]{|#1\rangle}

\newcommand{\qbinom}[3]{\genfrac{[}{]}{0pt}{}{\,#1\,}{#2}_{#3}}
\newcommand{\cspace}{$C^\infty(\mbb{R}^N)$~}
\newcommand{\tspace}{$C^\infty(\mbb{R}^N) \otimes \mc{S}$~}
\newcommand{\eq}[1]{(\ref{#1})}
\newcommand{\Eq}[1]{Eq.~(\ref{#1})}
\newcommand{\tr}{\operatorname{tr}}

\renewcommand{\mod}{\operatorname{mod}}

\begin{document}

\begin{center}
{\Large \bf \sf Partition functions of Polychronakos like spin chains 
associated with polarized spin reversal operators}

\vspace{1.3cm}

{\sf B. Basu-Mallick$^1$\footnote{
Corresponding author, Fax: +91-33-2337-4637, Telephone: +91-33-2337-5345, 
E-mail address:
bireswar.basumallick@saha.ac.in},
Nilanjan Bondyopadhaya$^2$\footnote{E-mail address:
nilanjan.iserc@visva-bharati.ac.in}
and Pratyay Banerjee$^1$\footnote{E-mail address: pratyay.banerjee@saha.ac.in}}
\bigskip

{\em $^1$Theory Division, Saha Institute of Nuclear Physics, \\
1/AF Bidhan Nagar, Kolkata 700 064, India}

\bigskip

{\em $^2$Integrated Science Education and Research Centre, \\ 
Siksha-Bhavana, Visva-Bharatai, Santiniketan 731 235, India}

\end{center}

\bigskip
\bigskip

\noi {\bf Abstract}
\vspace {.2 cm}

We construct polarized spin reversal operator (PSRO) which yields a class of 
representations for the $BC_N$ type of Weyl algebra, and subsequently use this 
PSRO to find out novel exactly solvable variants of the $BC_N$ type of spin 
Calogero model. The strong coupling limit of such spin Calogero models 
generates the $BC_N$ type of Polychronakos spin chains with PSRO. We derive 
the exact spectra of the $BC_N$ type of spin Calogero models with PSRO and 
compute the partition functions of the related spin chains by using the 
freezing trick. We also find out an interesting relation between the partition 
functions of the $BC_N$ type and $A_{N-1}$ type of Polychronakos spin chains.
Finally, we study spectral properties like level density and distribution 
of spacing between consecutive energy levels for $BC_N$ type of Polychronakos 
spin chains with PSRO. 

\vspace{.74 cm}
\noi PACS No. : 02.30.Ik, 05.30.-d, 75.10.Pq, 75.10.Jm 

\vspace {.2 cm}
\noindent Keywords: Exactly solvable quantum spin models,
   Polarized spin reversal operator, Partition function

\newpage

\baselineskip 16 pt
 
\noi \section{Introduction}
\renewcommand{\theequation}{1.{\arabic{equation}}}
\setcounter{equation}{0}
\medskip

Exactly solvable one-dimensional quantum many body systems with 
long-range interactions have been studied intensively during last few decades
\cite{Ca71,Su71,Su72, OP83, Ha88,Sh88,Po93,Po94,Fr93,FG05,BUW99,BB06} and 
have been applied in various topics of contemporary
physics as well as mathematics like generalized exclusion statistics,
electric transport in mesoscopic systems, ${\cal N}=4$ super Yang-Mills theory,
random matrix theory, 
multivariate orthogonal polynomials and Yangian quantum
groups \cite{Ha96,MS94,Po06,BR94,Ca95,BKS03,BBL09,Bea12,
TSA95,Fo94,UW97,BF97npb,BGHP93,Hi95npb,Ba99,HB00,BBHS07,BBS08}.  
Investigation of this type of quantum mechanical systems 
having only dynamical degrees of freedom was initiated by Calogero, who   
found the exact spectrum of a Hamiltonian describing particles  
 on a line, subject to a harmonic 
confining potential and two-body long-range interaction 
inversely proportional to the square of the  inter-particle
distances ~\cite{Ca71}. An exactly solvable 
 trigonometric variant of this rational Calogero model,
with particles moving on a circle and interacting through 
 two-body potentials proportional to the inverse square of their
chord distances, was subsequently studied by Sutherland~\cite{Su71,Su72}. 

In a parallel development, Haldane and Shastry pioneered
the study of quantum integrable spin chains with long-range interaction
~\cite{Ha88,Sh88}. 
They found an exactly solvable 
quantum spin-\half chain with long-range interactions, 
whose ground state coincides with the
$U\rightarrow \infty $ limit of Gutzwiller's variational wave function 
for the Hubbard model, and
yields a one-dimensional analogue
of the resonating valence bond state.
The lattice sites of this \su{2}
Haldane--Shastry (HS) spin chain are equally spaced on a circle and
all spins interact with each other
through pairwise exchange interactions inversely proportional 
to the square of their chord distances. Integrable 
models possessing  both spin and dynamical degrees of freedom,   
like \su{m} spin generalization of the Sutherland model, have 
been studied subsequently in the literature~\cite{HH92,HW93,MP93}.
Furthermore, a close connection between the  
\su{m} spin generalization of the Sutherland model
and the HS chain with \su{m} spin degrees of freedom has been   
established by using the method of `freezing trick'~\cite{Po93,SS93}. 
Indeed, by applying the above mentioned method, it can be shown 
that in the strong coupling limit the particles of the \su{m} spin 
Sutherland model `freeze' at
 the equilibrium position of the scalar part of the 
potential, and the dynamical
and spin degrees of freedom decouple from each other. Moreover, since
such equilibrium positions of the particles coincide with the equally spaced
lattice points of the HS spin chain, the dynamics of 
the decoupled spin degrees of freedom naturally leads to 
the Hamiltonian of the \su{m} HS model. In a similar way, 
application of this freezing trick to the 
\su{m} spin Calogero model with harmonic confining potential yields 
the Polychronakos spin chain 
(also known as Polychronakos--Frahm (PF) spin chain in the literature)
with Hamiltonian given by \cite{Po93,Fr93} 
\beq
\label{a1}
\mc{H}_\mr{PF}^{(m)}=\sum_{1\leq i< j \leq N}
\frac{1+P_{ij}}{(\rho_i-\rho_j)^2}  \, ,
\eeq
where $\rho_i$ denotes the $i$-th zero of the 
Hermite polynomial of degree $N$ and 
$P_{ij}$ is the exchange operator interchanging the `spins' 
(taking $m$ possible values) of $i$-th and $j$-th lattice sites. 
Thus, unlike the case of HS spin chain,  the lattice sites of the PF spin chain 
are inhomogeneously distributed on a line. 
Due to the decoupling of  
the spin and dynamical degrees of freedom of the \su{m} spin Calogero model
for large values of its coupling constant, an expression for  
the partition function of the \su{m} PF spin chain 
can be obtained by first computing the spectrum and
 partition function of the 
\su{m} spin Calogero model and then dividing such partition function 
by that of the spinless Calogero model~\cite{Po94}.
Similarly, the partition function of \su{m} HS spin chain 
can be computed by dividing the partition function of the 
\su{m} spin Sutherland model at the strong coupling limit by that of the 
spinless Sutherland model~\cite{FG05}.

The Hamiltonians of the above mentioned translational invariant 
\su{m} HS and PF spin chains, in which 
the strength of interaction between any two 
spins depends only on the difference of their site coordinates,  
have a close connection with the $A_{N-1}$ type of root system~\cite{OP83}. 
Variants of these spin chains associated with other root systems
have also been studied in the literature and applied 
in the context of one dimensional physical systems with boundaries which break 
the translational invariance. In particular, the spectrum of an equally 
spaced spin-\half HS chain related to 
the $BC_N$ root system has been 
studied by Bernerd et al.~\cite{BPS95}. 
A key feature in the Hamiltonian of this spin chain
is the presence of reflection operators like $\hat{P}_i$ (defined on the 
$i$-th lattice site) satisfying the relation  $\hat{P}_i^2=\one$.  
Since the internal space 
associated with each lattice site is two dimensional for this spin chain, 
the reflection operator yields 
three inequivalent representations: $\hat{P}_i=\pm \one $ 
and $\hat{P}_i=\sigma^x$,
where $\sigma^x$ is a Pauli matrix.  For the case  $\hat{P}_i= \one $ 
(or, $\hat{P}_i= -\one $), this spin-\half chain becomes \su{2}
invariant and coincides (up to an additive constant) 
with a spin model with open boundary condition,  
which was first considered by Simons and Altshuler \cite{SA94}.
On the other hand, for the case  $\hat{P}_i= \sigma^x$, where  
$\hat{P}_i$ can be interpreted as the spin reversal operator 
due to its action on the states of the $i$-th lattice site as  
$\hat{P}_i \v \frac{1}{2} \r = \v\!-\!\frac{1}{2}\r$,
$\hat{P}_i\v\!- \frac{1}{2} \!\r = \v  \frac{1}{2} \r$, 
this  spin-\half chain associated with the $BC_N$ root system 
breaks the \su{2} symmetry. 

 Taking $\hat{P}_i$ as the spin reversal operator (denoted by $P_i$) 
for any possible value of the `spin' degrees of freedom ($m\geq 2$),
and also allowing the possibility of having unequally spaced lattice sites
on a circle, the above mentioned 
HS spin chain associated with the $BC_N$ root system has been   
generalized by Enciso et al.~\cite{EFGR05}.
By employing the freezing trick,
the partition functions for this type of generalization 
of the HS spin chain and a similar generalization of the PF spin chain 
have also been calculated for all values of $m$~\cite{EFGR05,BFGR08}. 
However, to the best of our knowledge, 
 the partition functions for the Simons-Altshuler (SA)
type generalizations of HS and PF spin chains,
corresponding to the cases $\hat{P}_i= \pm \one $, 
 have not been computed till now for any value of $m$. Since SA type  
generalizations of HS and PF spin chains would be \su{m} invariant, 
exact solutions of these spin chains may play an important role in 
describing boundary effects in physical systems
which break the translational invariance but respect the 
internal \su{m} symmetry.

Even though $\hat{P}_i=\pm \one $ and
$\hat{P}_i = \sigma^x$ are the only possible inequivalent representations
of the reflection operator $\hat{P}_i$ for the case $m=2$, 
in this paper it will be shown that  
the situation is slightly more complex for the case $m>2$. 
Since each inequivalent representation of the 
reflection operator $\hat{P}_i$ on a complex $m$-dimensional vector space   
may lead to a different type of 
HS or PF spin chain associated with the $BC_N$ root system, 
at present our main aim is to construct all possible 
inequivalent representations of 
$\hat{P}_i$ for any value of $m$ and compute 
the partition functions of the corresponding PF spin chains 
through the freezing trick. Interestingly, it will turn out that, 
in general a representation of $\hat{P}_i$
can be characterized as a polarized spin reversal operator (PSRO)
which acts like the identity operator on some spin components and 
acts like the spin reversal operator  on the rest of the spin components.
In a particular limit, such PSRO
 coincides with the usual spin reversal operator $P_i$
which changes the signs of all spin components and, 
in the opposite limit,  
such PSRO yields $\hat{P}_i= \one$ (or, $\hat{P}_i= -\one$). 
The latter representation of $\hat{P}_i$ would allow us to construct
a \su{m} invariant SA type generalization of the PF spin chain, 
which is described by the Hamiltonian 
\beq
\label{a2}
{\mc{H}}^{(m,0)}= \sum_{1\leq i \neq j \leq N}
 \frac{y_i+y_j}{(y_i-y_j)^2} \, 
  (1+P_{ij})  \, ,
\eeq
where  $y_i$ denotes the $i$-th zero  
of the generalized Laguerre polynomial $L_N^{\beta -1}$.
Hence, the lattice sites of this \su{m} invariant Hamiltonian 
(\ref{a2}) implicitly depend on the real positive parameter $\beta$.
  
The organization of this paper is as follows. In Section 2, 
at first we review the key role played by the $BC_N$ type of Weyl algebra
in deriving the spectrum of the $BC_N$ type of spin Calogero model. 
Then we construct the PSRO which, along with the spin 
exchange operator $P_{ij}$, yields new representations of the $BC_N$ 
type of Weyl algebra in the internal space associated with
$N$ number of particles or lattice sites. 
In Section 3,  we use such PSRO to obtain novel exactly solvable
variants of the $BC_N$ type of spin Calogero model and 
subsequently take the strong coupling limit 
of these models to construct $BC_N$ type of PF spin
chains with PSRO. Next, we derive 
the exact spectrum of the $BC_N$ type of spin Calogero models with PSRO and 
also compute the partition functions of the related spin 
chains by using the freezing trick.
In Section 4, we derive an interesting relation between the partition 
function of the $BC_N$ type of PF spin chain with PSRO 
and that of the $A_{N-1}$ type of PF spin chain. Then we establish a  
duality relation between the partition functions 
of the $BC_N$ type of anti-ferromagnetic and ferromagnetic PF spin chains 
with PSRO. In Section 5, we compute the ground state and the highest 
state energy levels corresponding to the $BC_N$ type of PF spin 
chains with PSRO. In Section 6, we  study a few spectral  
properties of the $BC_N$ type of PF spin chains with PSRO,
like the energy level density and   
nearest neighbour spacing distribution.
In Section 7, we summarize our results and also 
mention some possible directions for future study. 

\bigskip

\noi \section{Construction of the PSRO}
\renewcommand{\theequation}{2.{\arabic{equation}}}
\setcounter{equation}{0}
\medskip

Similar to the case of $A_{N-1}$ type of quantum integrable 
systems with long-range interaction, $BC_N$ type of Dunkl operators 
and the corresponding auxiliary
operator (which is a quadratic sum of all Dunkl operators) 
play a central role in calculating the exact spectrum of the $BC_N$ type 
of spin Calogero model and its scalar counterpart~\cite{BFGR08}.
The form of such $BC_N$ type of auxiliary operator
is given by
\beq
\mbb{H}=-\sum_{i=1}^N \frac{\partial^2}{ \partial x_i^2 }
+a \, \sum_{i\neq j}  \left[
 \frac{a-K_{ij}}{(x_{ij}^-)^2}+ \frac{a-\widetilde{K}_{ij}}{(x_{ij}^+)^2}\right]
+\beta a \, \sum_{i=1}^N \frac{\beta a-K_i}{x_i^2}+\frac{a^2}4\,r^2\,,
\label{auxi}
\eeq
where $a> \frac{1}{2}, ~\beta>0$ are some real coupling constants and  
the notations $x_{ij}^- \equiv x_i-x_j$, $x_{ij}^+ \equiv x_i+x_j$
and $r^2\equiv \sum_{i=1}^N x_i^2$ are used. Moreover,  
$K_{ij}$ and $K_i$ are coordinate permutation and sign reversing
operators, defined by
\bse
\bea
&& (K_{ij}f)(x_1,\dots,x_i,\dots,x_j,\dots,x_N)=f(x_1,\dots,x_j,\dots,x_i,
\dots,x_N)\,,  \label{action1}\\
&& (K_i f)(x_1,\dots,x_i,\dots,x_N)=f(x_1,\dots,-x_i,\dots,x_N)\,, 
\label{action2} 
\eea 
\label{action}
\ese
and $\widetilde{K}_{ij}=K_iK_jK_{ij}$. Thus the operators 
$K_{ij}$, $K_i$ and $\mbb{H}$ act on the functions of the coordinate space, 
which is denoted by \cspace. By using \Eq{action}, it is easy to check that 
 $K_{ij}$ and $K_i$  give a realization of the  $BC_N$ type of 
Weyl algebra generated by $\mc{W}_{ij}$ and $\mc{W}_{i}\,$:
\begin{subequations}
\bea
&\mc{W}_{ij}^2=\one \, , ~~~~\mc{W}_{ij}\mc{W}_{jk} =\mc{W}_{ik}\mc{W}_{ij}
=\mc{W}_{jk}\mc{W}_{ik} \, ,~~~~ \mc{W}_{ij}\mc{W}_{kl}
=\mc{W}_{kl}\mc{W}_{ij} \, ,  \label{weyl1}\\ 
&\mc{W}_{i}^2=\one \, , ~~~~ \mc{W}_{i}\mc{W}_{j}= \mc{W}_{j}\mc{W}_{i} \, ,
~~~~ \mc{W}_{ij}\mc{W}_{k}=\mc{W}_{k}\mc{W}_{ij} \, , 
~~~~\mc{W}_{ij}\mc{W}_{j}=\mc{W}_{i}\mc{W}_{ij} \, . \label{weyl2}
\eea
\label{weyl}
\end{subequations}
The Hamiltonian of the $BC_N$ type of spin Calogero model, 
as considered in Ref.~\cite{BFGR08}, is quite 
similar in form to that of the auxiliary operator \eq{auxi}.
However, this Hamiltonian acts 
not only on the functions of the coordinate space, but on a 
direct product space like 
\tspace, where
\beq
\mc{S} \equiv \underbrace{\mc{C}_m \otimes \mc{C}_m \cdots 
\otimes \mc{C}_m}_{N} \, ,
\label{spinspace}
\eeq
with $\mc{C}_m$ denoting the $m$-dimensional complex vector space 
associated with each particle. In terms of orthonormal basis vectors, 
the total spin space $\mc{S}$ may be expressed as     
\beq
\mc{S}= \Big{\langle} \ket{s_1,\cdots,s_N}^* ~  \Big{\vert} 
s_i\in \{-M,-M+1, \cdots , M \}; ~ M= \frac{m-1}{2} \Big{\rangle} \, . 
\label{basis}
\eeq
The spin exchange operator $P_{ij}$ and the spin reversal operator $P_i$
are defined on the space $\mc{S}$ as  
\begin{subequations}
\bea
&&P_{ij}\ket{s_1 \, , \cdots, s_i \, , \cdots, s_j \, , \cdots, s_N}^*
= 
\ket{s_1 \, , \cdots, s_j \, , \cdots, s_i \, , \cdots, s_N}^* \label{spin1} \\
&&P_{i}\ket{s_1 \, , \cdots, s_i \, , \cdots, s_N}^*
= \ket{s_1 \, , \cdots, - s_i \, , \cdots, s_N}^* \, .
 \label{spin2}
\eea
\label{spin}
\end{subequations}
It is easy to check that, similar to the case of $K_{ij}$ and $K_i$, 
 $P_{ij}$ and $P_i$ also give a realization of the  $BC_N$ type of 
Weyl algebra \eq{weyl}.
By using the operators $P_{ij}$ and $P_i$, one can define 
the Hamiltonian of the $BC_N$ type of spin Calogero model   
as~\cite{BFGR08}
\beq
H^{(m)}=-\sum_{i=1}^{N}\frac{\d^2}{\d x_i^2}+a\sum_{i\neq j}
\left[\frac{a+P_{ij}}{(x_{ij}^-)^2} +
\frac{a+\widetilde{P}_{ij}}{(x_{ij}^+)^2}\right]
+\beta a\sum_{i=1}^{N}\frac{\beta a-\ep P_i}{x_i^2}+\frac{a^2}{4}r^2  \, ,
\label{ham}
\eeq
where $\ep=\pm 1$ and $\wt{P}_{ij} \equiv P_iP_jP_{ij}$. 
Note that the Hamiltonian \eq{ham} of $BC_N$ spin Calogero model 
can be reproduced from the auxiliary operator \eq{auxi} through 
simple substitutions like 
\beq
H^{(m)}= \mbb{H}\big{\vert}_{K_{ij} \rightarrow -P_{ij},
\, K_i \rightarrow \ep P_i } \, . 
\label{subs}
\eeq
Consequently, the Hilbert space and the spectrum of  $H^{(m)}$ 
can be obtained from those of $\mbb{H}$ by applying a projector
$\Lambda$ which satisfies the relations~\cite{BFGR08}
\bse 
\bea
&&K_{ij}P_{ij}\Lambda=\Lambda K_{ij}P_{ij}=-\Lambda \label{pro1} \\
&& K_iP_i\Lambda=  \Lambda K_i \, P_i=  \ep\Lambda \label{pro2} \, . 
\eea  
\label{pro}
\ese 
For constructing the projector $\Lambda$, it is important to 
observe that both of the two 
sets of operators given by $K_{ij}, ~K_i$ and $P_{ij}, ~P_i$
yield realizations of the $BC_N$ type of Weyl algebra \eq{weyl} on the spaces 
\cspace and $\mc{S}$ respectively. Hence, it is possible to define another 
set of operators like
$\Pi_{ij}=K_{ij}P_{ij}, ~\Pi_{i}=K_iP_i$, which will yield 
a realization of the $BC_N$ type of Weyl algebra \eq{weyl} on the space 
\tspace. Let us now define an operator $\Lambda_0$ on the space 
\tspace as
\beq
\Lambda_0=\frac{1}{N!}\sum_{i=1}^{N!}\, \vep_l \, \mc{P}_l \, ,
\label{pro3}
\eeq
where $\mc{P}_l$ denotes an element of the realization
of the permutation group generated by the operators $\Pi_{ij}$
and $\vep_l$ is the signature of $\mc{P}_l$.
For example, corresponding to the simplest $N=2$ and $N=3$
cases, $\Lambda_0$ is given by
\bea
&&N=2: ~~~~~~ \Lambda_0  = \frac{1}{2}\left( 1- \Pi_{12}\right) \, ,
\nn \\
&&N=3: ~~~~~~ \Lambda_0  = \frac{1}{6}\left( 1- \Pi_{12}
-\Pi_{13}-\Pi_{23}+ \Pi_{12}\Pi_{13}+\Pi_{12}\Pi_{23}
\right) \, .
\nn
\eea
It is easy to show that $\Lambda_0$  in \Eq{pro3} 
satisfies the relations
\beq
\Lambda_0^2 = \Lambda_0,~~ K_{ij}P_{ij}\Lambda_0=\Lambda_0 K_{ij}P_{ij}
=-\Lambda_0 \, .
\nn
\eeq
 Hence $\Lambda_0$ acts as an antisymmetriser with respect  to the simultaneous 
interchange of the coordinate and the spin degrees of freedom.
With the help of this $\Lambda_0$, it is possible to finally construct 
the projector $\La$ as \cite{FGGRZ03}
\beq
\La = \frac{1}{2^N}\left(\prod_{j=1}^N\left(1+\ep\, \Pi_j\right)
\right)\La_0 \, .
\label{pro4}
\eeq
Using the fact that $\Pi_{ij}$ and $\Pi_{i}$ yield 
a realization of the $BC_N$ type of Weyl algebra \eq{weyl}, one can easily 
verify that the projector $\La$ satisfies the relations 
\eq{pro}. Hence, with the help of the projector given in \eq{pro4}, 
it is possible to compute the spectrum of  $H^{(m)}$ from the 
known spectrum of the auxiliary operator.

Even though the projector \eq{pro4} is constructed for a  
particular representation of the $BC_N$ type of Weyl algebra \eq{weyl},
such a projector can also be written in an abstract algebraic form \cite{CC04}. 
Therefore, if we can modify the action of $P_i$ given in \Eq{spin2} so that, 
along with $P_{ij}$ in \eq{spin1}, this modified version of $P_{i}$ 
yields an inequivalent representation of the $BC_N$ type of Weyl algebra 
\eq{weyl}, then it would be possible to explicitly construct 
the corresponding projector in exactly same way.  
Consequently, such modified version 
of $P_i$ would lead to a new $BC_N$ type of spin Calogero model 
whose spectrum can be computed by using the method of projector.
For the purpose of finding out modified versions of $P_i$ which may give   
inequivalent representations of the $BC_N$ type of Weyl algebra, 
at first we notice that the spin reversal operator $P_i$ in \Eq{spin2}
acts nontrivially only on the $i$-th spin space. Hence, 
this $P_i$ can also be written in the form
\beq
\begin{array}{clc}
{P_i = \one \otimes \cdots \otimes \one} & {\otimes ~~ P ~~\otimes} &
 {\! \! \one \otimes \cdots \otimes \one}  \, , \\  
{}&{\text{i-th place}}&{}  
\end{array}
\label{product}
\eeq
where $P$ acts on $\mc{C}^m$ as 
\beq
P\ket{s_i}^*=\ket{-s_i}^* \, .
\label{reve}
\eeq
In analogy with this case, we assume that all modified versions of $P_i$ 
act nontrivially only on the $i$-th spin space.
Due to the relation $\mc{W}_{i}^2=\one$ within~\Eq{weyl2}, such 
modified versions of $P_i$ can be treated as involutions 
on the $i$-th spin space. It is known that 
the Hamiltonian of the $BC_N$ type of spin Calogero model, 
with reflection operators formally  
defined as involutions on the corresponding spin spaces, 
yields a quantum integrable system 
with mutually commuting conserved quantities 
\cite{YT96}. Consequently, the spin Calogero models which 
we shall construct in the next section by using 
modified versions of $P_i$ would also represent quantum
integrable systems. 

For the purpose of explicitly finding out all possible 
modified versions of $P_i$, which act as involutions on the 
$i$-th spin space and also satisfy the 
$BC_N$ type of Weyl algebra \eq{weyl}, 
let us  arbitrarily partition $m$  into two parts as $m=m_1+m_2$, 
where $m_1\geq m_2 \geq 0$.
Evidently, the internal space $\mc{C}_m$ associated with the $i$-th particle 
can always be written as a direct sum of any two orthogonal subspaces 
of dimension $m_1-m_2$ and $2m_2$ respectively:
\beq
\mc{C}_m= \mc{C}_{m_1-m_2} \oplus \mc{C}_{2m_2} \, ,
\label{dsum}
\eeq
where  $\mc{C}_{m_1-m_2}$ and $\mc{C}_{2m_2}$ are defined in terms of 
orthonormal basis vectors as 
\beq
\mc{C}_{m_1-m_2}= \Big{\langle} \ket{\alpha}' ~  \Big{\vert} 
\, \alpha \in \{1,2, \cdots , m_1-m_2 \}\Big{\rangle} \, ,~
\mc{C}_{2m_2}= \Big{\langle} \ket{\beta}'' ~  \Big{\vert} 
\, \beta \in \{ 1,2, \cdots , 2m_2 \} \Big{\rangle} \, . 
\label{basis2}
\eeq
In analogy with \Eq{product}, we propose a modification of $P_i$ 
in the form 
\beq
\begin{array}{clc}
{P_i = \one \otimes \cdots \otimes \one} & {\otimes ~~ 
P^{(m_1,m_2)} ~~\otimes} &
 {\! \! \one \otimes \cdots \otimes \one}  \, , \\  
{}&{\text{~~~~i-th place}}&{}  
\end{array}
\label{product1}
\eeq
where $P^{(m_1,m_2)}$ acts in a rather different way on the two subspaces  
 $\mc{C}_{m_1-m_2}$ and $\mc{C}_{2m_2}$
of the space $\mc{C}_m$. More precisely, $P^{(m_1,m_2)}$
acts like an identity operator 
on the space $\mc{C}_{m_1-m_2}$ and acts like
a spin reversal operator on the even dimensional space  
$\mc{C}_{2m_2}$. Thus, the action of $P^{(m_1,m_2)}$ 
 on the basis vectors of $\mc{C}_{m_1-m_2}$ is given by 
\beq
P^{(m_1,m_2)} \ket{\alpha}'= \ket{\alpha}'\, .
\label{iden1}  
\eeq
Moreover, in analogy with \Eq{reve}, the action of $P^{(m_1,m_2)}$ 
on the first basis vector of $\mc{C}_{2m_2}$ would give the 
last basis vector of this space, on the second basis vector 
would give the last but one basis vector, and so on.
Hence, in general, the action of $P^{(m_1,m_2)}$ 
on the basis vectors of $\mc{C}_{2m_2}$
may be written as  
\beq
P^{(m_1,m_2)} \ket{\beta}'' = \ket{2m_2+1-\beta}'' \, . 
 \label{refl}
\eeq
Since $P^{(m_1,m_2)}$
acts like a spin reversal operator only on a subspace of $\mc{C}_{m}$,
and acts trivially on the complementary subspace, 
it is natural to call 
$P_i^{(m_1,m_2)}$ as a PSRO associated with the $i$-th particle.
Note that the relation  
 $\left(P^{(m_1,m_2)}\right)^2= \one$ is satisfied for both of the spaces 
$\mc{C}_{m_1-m_2}$ and $\mc{C}_{2m_2}$.  
For the purpose of representing $P^{(m_1,m_2)}$ in a more convenient form,
let us take another set of orthonormal basis vectors of $\mc{C}_{2m_2}$ as
\beq
\ket{\be}_{\pm}=\frac{1}{\sqrt{2}} \, 
\big( \, \ket{\be}'' \pm \ket{2m_2+1-\be}'' \, \big) \, ,
\label{newb}
\eeq 
where $\be\in\{1, 2, \ldots, m_2\}$. 
By using \Eq{refl}, it is easy to check that 
\beq
P^{(m_1,m_2)}\ket{\be}_{\pm}= \pm \, \ket{\be}_{\pm} \, .
\label{diag}
\eeq 
Due to \Eq{dsum}, we can choose an orthonormal set of basis 
vectors for the space $\mc{C}_{m}$ as
\beq
\mc{C}_{m} = \Big{\langle} \, \ket{s}  \, \Big{\vert} 
\, s \in \{1,2, \cdots , m_1+m_2 \} \, 
\Big{\rangle} \, , 
\label{basis3} 
\eeq
where $\ket{s}=\ket{\al}'$ with $\al=s$ for 
$s \in \{1,2, \cdots , m_1-m_2 \}$,
$\ket{s}=\ket{\be}_+$ with $\be=s - m_1 + m_2$ for 
$s \in \{m_1-m_2+1,m_1-m_2+2, \cdots , m_1 \}$ and 
$\ket{s}=\ket{\be}_-$ with $\be= s -m_1$ for 
$s \in \{m_1+1,m_1+2, \cdots , m_1+m_2 \}$.
Using Eqs.~\eq{iden1} and \eq{diag}, it is easy to show that  
$P^{(m_1,m_2)}$ acts as a diagonal matrix on the basis vectors \eq{basis3}
of $\mc{C}_{m}$:
\beq
P^{(m_1,m_2)} = 
\begin{pmatrix}
1& \\
&\ddots & \\
&& 1 & \\
&&&-1 & \\
&&&&\ddots & \\
&&&&&-1 & \\
\end{pmatrix} \,, 
\label{diag1}
\eeq
where there are $m_1$ number of 1's and $m_2$ number of -1's 
along the main diagonal. Combining Eqs.~\eq{spinspace} and \eq{basis3},
we express the total spin space $\mc{S}$ through a set of
orthonormal basis vectors as 
\beq
\mc{S}= \Big{\langle} \ket{s_1,\cdots,s_N} ~  \Big{\vert} 
s_i \in \{ 1,2, \cdots , m \} \Big{\rangle} \,  . 
\label{basis1}
\eeq
Due to Eqs.~\eq{product1} and \eq{diag1}, 
$P_i^{(m_1,m_2)}$ acts on these basis vectors as 
\beq
P_i^{(m_1, m_2)}\ket{s_1, \cdots, s_i,\cdots , s_N}
=(-1)^{f(s_i)}\ket{s_1, \cdots, s_i,\cdots , s_N},
\label{diag2}
\eeq 
where
\begin{equation}
 f(s_i) = \left \{  
\begin{array}{ll} 
0,  & \mbox{~for $s_i\in\{1, 2, \ldots, m_1\}$, } \\
1, &   \mbox{~for $s_i\in\{m_1+1, \ldots, m_1+m_2\}$. } 
\end{array} 
\right. \nn
\end{equation}
In analogy with \Eq{spin1}, we define the action of $P_{ij}$
on the basis vectors \eq{basis1} as 
\beq
P_{ij}\ket{s_1 \, , \cdots, s_i \, , \cdots, s_j \, , \cdots, s_N}
= 
\ket{s_1 \, , \cdots, s_j \, , \cdots, s_i \, , \cdots, s_N} \, . 
\label{spin4} 
\eeq
Using Eqs.~\eq{diag2} and \eq{spin4},
one can easily check that $P_i^{(m_1,m_2)}$
and $P_{ij}$ yield a realization of the  $BC_N$ type of 
Weyl algebra \eq{weyl}. 
In this context it may be recalled that, while  
constructing $P_i^{(m_1,m_2)}$ as a PSRO, 
we have previously assumed that $m_1\geq m_2$.
However, this condition is really not necessary 
for showing that
$P_i^{(m_1,m_2)}$ and $P_{ij}$ yield a 
realization of the $BC_N$ type of Weyl algebra. Therefore, in the rest
of this article we shall take \Eq{diag2}, with any possible values
of $m_1$ and $m_2$ satisfying the condition $m_1+m_2=m$,
as the definition of  PSRO. Since the  
trace of $P_i^{(m_1,m_2)}$ in \Eq{diag2} is given by 
\beq
 \tr P_i^{(m_1,m_2)} = m^{N-1}(m_1-m_2) \, ,
 \label{trace1}
\eeq
it is evident that, for any given value of $m$, $P_i^{(m_1,m_2)}$
with each distinct set of values for $m_1$ and $m_2$ 
would lead to an inequivalent realization of the $BC_N$ type of Weyl algebra.
In the next section, we shall use such PSRO
 to obtain new exactly solvable
variants of the $BC_N$ type of spin Calogero model \eq{ham} and the  
related PF spin chain. 
 It may be observed that the trace of
the spin reversal operator $P_i$ in \Eq{spin2} is given by
\beq
 \tr P_i = 
m^{N-1} \times \left(m \mod 2
\right) \, .
 \label{trace2}
\eeq
Comparing \Eq{trace1} with \Eq{trace2} we find that, 
the trace of $P_i^{(m_1,m_2)}$ coincides with that of
$\ep P_i$ in the special case 
$m_1=m_2$ ($m_1=m_2+\ep$) for even (odd) values of $m$.
Since both of the operators
 $P_i^{(m_1,m_2)}$ and $\ep P_i$ can only have eigenvalues $\pm 1$,   
these two operators yield exactly same set of eigenvalues and lead to 
equivalent representations of the $BC_N$ type of Weyl algebra
 for the above mentioned choice of $m_1$ and $m_2$.
It may also be noted that, for the special case $ m_1=m,~ m_2=0$,  
$P_i^{(m_1,m_2)}$ in \Eq{diag2} reduces to the trivial identity
operator.

\bigskip

\noi \section{Spectra and partition functions of $BC_N$ type models with PSRO }
\renewcommand{\theequation}{3.{\arabic{equation}}}
\setcounter{equation}{0}
\medskip
In this section, we shall use the PSRO
 for obtaining new variants of the $BC_N$ type of spin Calogero model \eq{ham}
and subsequently take the strong coupling limit 
of such spin Calogero models to construct 
the corresponding $BC_N$ type of PF spin chains. 
Next, by using the method of projector which has been discussed 
in the previous section, we shall find out 
the exact spectrum of $BC_N$ type of spin Calogero models with PSRO.
Finally we shall compute the partition functions of the $BC_N$ type of PF spin 
chains with PSRO by using the freezing trick.

Substituting $\ep P_i$ by $P_i^{(m_1,m_2)}$
in the Hamiltonian \eq{ham}, we obtain the Hamiltonians of 
the $BC_N$ type of spin Calogero models with PSRO as 
 \beq
H^{(m_1,m_2)}=-\sum_{i=1}^{N}\frac{\d^2}{\d x_i^2}+a\sum_{i\neq j}
\left[\frac{a+P_{ij}}{(x_{ij}^-)^2} +
\frac{a+\widetilde{P}_{ij}^{(m_1,m_2)}}{(x_{ij}^+)^2}\right]
+\beta a\sum_{i=1}^{N}\frac{\beta a -P_i^{(m_1,m_2)}}{x_i^2}+\frac{a^2}{4}r^2 ,
\label{A2}
\eeq
where $\widetilde{P}_{ij}^{(m_1,m_2)} \equiv 
 P_i^{(m_1,m_2)}P_j^{(m_1,m_2)} P_{ij}$. 
Since $P_i^{(m_1,m_2)}$ and $\ep P_i$ yield  
equivalent representations of the $BC_N$ type of Weyl algebra
in the special case 
$m_1=m_2$ ($m_1=m_2+\ep$) for even (odd) values of $m$, 
 $H^{(m_1,m_2)}$ \eq{A2} would reduce to $H^{(m)}$ \eq{ham} 
after an appropriate  
similarity transformation in this special case. 
In another special case  given by $m_1=m,~ m_2=0$,  
where $P_i^{(m_1,m_2)}$ reduces to the identity
operator, $H^{(m_1,m_2)}$ \eq{A2} yields an 
SA type extension of spin Calogero model given by  
\beq
H^{(m,0)}=-\sum_{i=1}^{N}\frac{\d^2}{\d x_i^2}+a\sum_{i\neq j}
(a+P_{ij})
\left[\frac{1}{(x_{ij}^-)^2} +
\frac{1}{(x_{ij}^+)^2}\right]
+\beta a(\beta a-1)\sum_{i=1}^{N}\frac{1}{x_i^2}+\frac{a^2}{4}r^2 \, .
\label{A2a}
\eeq
It may be noted that, the above Hamiltonian has been obtained  
earlier by using an auxiliary operator, which was constructed  
 through a combination of several $A_{N-1}$ type of Dunkl operators 
\cite{FGGRZ01}. However, at present we have obtained this Hamiltonian
\eq{A2a} as a special case of \eq{A2}, which will be shown to be 
related to the $B_N$ type of Dunkl operators.
Thus the Hamiltonian $H^{(m,0)}$ is surprisingly related 
to both  $A_{N-1}$ and $B_N$ types of Dunkl operators. 

Since the potentials of the Hamiltonian $H^{(m_1,m_2)}$ \eq{A2}
become singular in the limits $x_i \pm x_j \to 0$ and 
$x_i \to 0$, the configuration space of this Hamiltonian can be 
taken as one of the maximal open subsets of $\mbb{R}^N$ on which 
linear functionals $x_i\pm x_j$ and $x_i$ have constant signs, 
i.e., one of the Weyl chambers of the $BC_N$ root system. 
Let us choose this configuration space as the principal 
Weyl chamber given by 
\beq
 C = \{ \mbf{x} \equiv (x_1,x_2,\cdots ,x_N) :~ 0<x_1<x_2<\ldots<x_N \} \, .
\label{conf}
\eeq 
Note that this configuration 
space does not depend on the values of $m_1$ and $m_2$, 
and coincides with the configuration space of $H^{(m)}$ \cite{BFGR08}. 
The Hamiltonian of the $BC_N$ type of PF spin chains with PSRO can be 
obtained from the Hamiltonian \eq{A2} in the limit $a\to \infty $ 
by means of the freezing trick. To this end, 
we express $H^{(m_1,m_2)}$ \eq{A2} 
in powers of the coupling constant $a$ as 
\beq
H^{(m_1,m_2)} = -\sum_{i=1}^N \frac{\partial^2}{ \partial x_i^2 }
+ a^2 \, U + O(a) \, ,
\label{order}
\eeq 
with 
\beq
U(x)=\sum_{i\neq j}\left[\frac{1}{(x_{ij}^-)^2}+\frac{1}{(x_{ij}^+)^2}\right]
+\beta^2\sum_{i=1}^{N}\frac{1}{x_i^2}+\frac{r^2}{4}.
\label{A5}
\eeq
As the coefficient of $a^2$ order term in \eq{order} dominates
 in the limit $a\to \infty $, the particles 
of the spin dynamical model \eq{A2} concentrate at the coordinates 
$\xi_i$ of the minimum $\mbs{\xi}$ of the potential $U$ in $C$.
Since the Hamiltonian \eq{A2}  can be written
in the form  
\beq
H^{(m_1,m_2)}=H_{sc}+a\, \mathfrak{H}^{(m_1,m_2)} \, ,
\label{A3}
\eeq
where $H_{sc}$ is the scalar (spinless) Calogero model of $B_N$ type given by 
\beq
H_{sc}=-\sum_{i=1}^N \frac{\partial^2}{\partial x_i^2}+a(a-1)\sum_{i\neq 
j}\bigg[
\frac{1}{(x_{ij}^-)^2}+\frac{1}{(x_{ij}^+)^2}\bigg]
+\sum_i\frac{a\beta (a \beta -1)}{x_i^2}+\frac{a^2}{4}\,r^2\,,
\label{A5a}
\eeq 
and 
\beq
\mathfrak{H}^{(m_1,m_2)}
=\sum_{i\neq j}\left[\frac{1+P_{ij}}{(x_i-x_j)^2} +
\frac{1+\widetilde{P}_{ij}^{(m_1,m_2)}}{(x_i+x_j)^2}\right]
+\beta\sum_{i=1}^{N}\frac{1-P_i^{(m_1,m_2)}}{x_i^2} \, ,
\label{A4a}
\eeq
it follows that the dynamical and internal degrees of freedom 
of $H^{(m_1,m_2)}$
decouple from each other in the limit $a\to \infty$. Moreover,
in this freezing limit, 
 the internal degrees of freedom of 
$H^{(m_1,m_2)}$ are governed by the Hamiltonian 
$\mathcal{H}^{(m_1,m_2)} =
\mathfrak{H}^{(m_1,m_2)}|_{\mbf{x} \to \mbs{\xi}}$, which is 
explicitly given by 
\beq
\mathcal{H}^{(m_1,m_2)}
=\sum_{i\neq j}\left[\frac{1+P_{ij}}{(\xi_i-\xi_j)^2} +
\frac{1+\widetilde{P}_{ij}^{(m_1,m_2)}}{(\xi_i+\xi_j)^2}\right]
+\beta\sum_{i=1}^{N}\frac{1-P_i^{(m_1,m_2)}}{\xi_i^2} \, . 
\label{A4}
\eeq
The operator $\mathcal{H}^{(m_1,m_2)}$
in the above equation represents the Hamiltonian of the    
$BC_N$ type of PF spin chain with PSRO, whose lattice sites $\xi_i$ are 
the coordinates of the unique minimum $\mbs{\xi}$ 
of the potential $U$ \eq{A5} within the configuration space $C$ \eq{conf}.
The uniqueness of such minimum was established in Ref.~\cite{CS02}
by expressing the potential $U$
in terms of the logarithm of the ground state wave function  
of the scaler Calogero model \eq{A5a}.   
The ground state wave function  
of this scaler Calogero model takes the form 
\beq
\mu({\bf x})=e^{-\frac{a}{4}r^2}~\prod_i|x_i|^{\beta a}
~\prod_{i<j}|x_i^2-x_j^2|^a,
\label{A11}
\eeq
and the corresponding ground state energy is given by 
\begin{equation}\label{E0}
E_0=Na\Big(\beta a +a(N-1)+\frac{1}{2} \Big)\,.
\end{equation}
Since the sites $\xi_i$ coincide with the coordinates of the 
(unique) critical point of $\log \mu({\bf x})$ in $C$, 
they can be determined through the set of relations  \cite{CS02,BFGR08}
\beq
\sum_{\stackrel{j=1}{(j\neq i)}}^N \, \frac{2y_i}{y_i-y_j} \, = \, y_i-\be \, ,
\label{glag}
\eeq
where $\xi_i=\sqrt{2y_i}$ and $y_i$'s satisfying \eq{glag} 
represent the zero points  
of the generalized Laguerre polynomial $L_N^{\beta -1}$.
Due to the presence of the operator $P_i^{(m_1,m_2)}$, 
the Hamiltonian \eq{A4} is not \su{m} invariant in general.
However, in the special case  given by $m_1=m,~ m_2=0$,  
$\mathcal{H}^{(m_1,m_2)}$ in \eq{A4} 
reduces to the \su{m} invariant
SA type generalization of the PF spin chain \eq{a2}, whose partition
function has not been computed till now.  
On the other hand, using a similarity 
transformation in the special case given by 
$m_1=m_2$ ($m_1=m_2+\ep$) for even (odd) values of $m$, 
$\mathcal{H}^{(m_1,m_2)}$ can be reduced to the Hamiltonian 
\beq
\mathcal{H}^{(m)}
=\sum_{i\neq j}\left[\frac{1+P_{ij}}{(\xi_i-\xi_j)^2} +
\frac{1+\widetilde{P}_{ij}}{(\xi_i+\xi_j)^2}\right]
+\beta\sum_{i=1}^{N}\frac{1-\ep P_i}{\xi_i^2} \, , 
\label{A4b}
\eeq
whose partition function has been computed earlier by using the 
freezing trick \cite{BFGR08}.

 
We have already seen that
 the spin and dynamical degrees of freedom of the Hamiltonian 
\eq{A2} decouple in the freezing limit $a\to \infty$. Hence, due to \Eq{A3}, 
eigenvalues of ${H}^{(m_1,m_2)}$ are approximately given by 
\beq
E_{ij} \simeq E_i^{sc} + a \, \mc{E}_j \, ,
\label{deco}
\eeq
where $E_i^{sc}$ and $\mc{E}_j$ are two arbitrary eigenvalues
of $H_{sc}$ and $\mathcal{H}^{(m_1,m_2)}$ respectively. 
By using the asymptotic relation \eq{deco}, one can easily  
derive an exact formula for the partition function $\mc{Z}_N^{(m_1,m_2)}(T)$
of the spin chain \eq{A4} as
\beq
\mc{Z}_N^{(m_1,m_2)}(T)=\lim_{a \rightarrow \infty}
 \frac{Z^{(m_1,m_2)}_N(aT)}{Z_N(aT)} \, ,
\label{A6}
\eeq
where $Z_N^{(m_1,m_2)}(aT)$ denotes the partition function 
for the spin dynamical model (\ref{A2}) and
$Z_N(aT)$ denotes that of the scalar model (\ref{A5a}).
Therefore, we can  evaluate the partition function 
$\mc{Z}_N^{(m_1,m_2)}(T)$ of the spin chain \eq{A4} 
 by computing first the spectra and partition functions of the 
Hamiltonians $H^{(m_1,m_2)}$ and $H_{sc}$. To this end, we shall
follow the approach of Ref.~\cite{BFGR08}, where the auxiliary
operator \eq{auxi} and related Dunkl operators 
have played a key role.
The form of rational Dunkl operators of $BC_N$ type are given by  
\bea
J_i^-=\frac{\partial}{\partial x_i} +a \sum_{j\neq i}\left[ \frac{1}{x_{ij}^-}
(1-K_{ij})\,
+\frac{1}{x_{ij}^+}(1-\widetilde{K}_{ij})\right] 
+\beta a\, \frac{1}{x_i}(1-K_i)\, ,
\label{J-}
\eea
where $i \in \{ 1,2, \dots,N \}$. The auxiliary operator \eq{auxi}
can be written through these Dunkl operators as 
\beq
\mathbb{H}= \mu(\mbf{x}) \left[ -\sum_i  \big(J_i^-\big)^2+a \sum_i x_i 
\frac{\partial}{\partial x_i}
+ E_0 \right] \mu^{-1}(\mbf{x})\, . 
\label{HJ}
\eeq
Evidently, the Dunkl operators~\eq{J-} map any monomial 
$\prod_i x_i^{n_i}$ into a
polynomial of total degree $n_1+ n_2+\cdots+n_N-1$. 
Therefore, if we consider a Hilbert space having a set of
 basis vectors like 
\beq
\phi_{\bf{n}}({\bf x})=\mu({\bf x})\prod_ix_i^{n_i} ,
\label{A10}
\eeq
with $n_i$'s being arbitrary non-negative integers,  
and partially order these basis 
vectors according to the total degree
$|\mbf{n}| \equiv n_1+n_2+\cdots + n_N $, 
then it follows from \Eq{HJ}  that  the operator
$\mathbb{H}$  would become an upper triangular matrix in the aforesaid
nonorthonormal basis. More precisely,
\beq
\mathbb{H} \phi_{\bf{n}}({\bf x})=E_{\mbf{n}} \phi_{\bf{n}}({\bf x})+
\sum_{\vert { \bf m} \vert < \vert {\bf
n} \vert }c_{{\bf m} {\bf n} }\, \phi_{\bf m} ({\bf x})\,,
\label{tria}
\eeq
where
\begin{equation}\label{Ep}
E_{\bf n}=a \vert  {\bf n}  \vert+E_0 \,  ,
\end{equation}
and the coefficients $c_{{\bf m} {\bf n} }$ are real constants.
Since the diagonal elements of any upper triangular matrix coincide with  
its eigenvalues, the spectrum of $\mbb{H}$ is given by \Eq{Ep}  
where $n_i$'s can be taken as arbitrary non-negative integers.

Note that the Hamiltonians of both scalar Calogero model (\ref{A5a}) 
and the spin Calogero model with PSRO (\ref{A2}) may be obtained from 
the auxiliary operator \eq{auxi} through formal  substitutions like 
\begin{subequations}
\bea
&&H_{sc}= \mathbb{H} |_{K_{ij},\, K_i \rightarrow 1} \, , \label{subs1} \\
&&H^{(m_1,m_2)}= \mathbb{H}|_{K_{ij} \rightarrow -P_{ij}^{(m_1,m_2)},\, 
K_i \rightarrow P_i^{(m_1,m_2)}} \,  .
\label{subs2} 
\eea
\label{subs}
\end{subequations}
Consequently, it would be possible to compute the spectra of these 
Hamiltonians from the known spectrum of the auxiliary operator 
with the help of appropriate projectors. 
For the purpose of obtaining the spectrum of $H_{sc}$ \eq{A5a}, 
one considers scalar functions of the form \cite{BFGR08}
\beq
\psi_{\mbf{n}}(\mbf{x})= \La_{sc} \, \phi_{\bf{n}}({\bf x}) \, ,
\label{swave}
\eeq
where $\La_{sc}$ is the symmetriser with respect to both permutations
and sign reversals, i.e., it satisfies the relations given by   
\beq
K_{ij}  \La_{sc}= \La_{sc}K_{ij}= \La_{sc} \, , ~~~~
K_{i}  \La_{sc}= \La_{sc}K_{i}= \La_{sc} \, . 
\label{spro}
\eeq
By using these relations, it can be shown that the functions 
\eq{swave} form a (nonorthonormal) basis of the Hilbert space 
of $H_{sc}$, provided that $n_i=2k_i$ are even integers and 
$k_1 \geq k_2 \geq \cdots \geq k_N$. As before, one can  
define a partial ordering among these basis vectors 
by comparing their degree. 
Due to Eqs.~\eq{subs1} and \eq{spro}, it follows that $H_{sc}$ \eq{A5a}
can be written as an upper triangular matrix with diagonal
elements $E^{sc}_{\bf n}$ also given by the right hand side 
of \Eq{Ep}. Thus one obtains 
 the exact partition function of the $BC_N$ type of  
scalar Calogero model (\ref{A5a}) as \cite{BFGR08}
\beq
Z_N(aT)=
\sum_{k_1 \geq k_2 \geq\dots \geq \,k_N \geq \, 0} q^{2 |{\bf k}|+\tilde{E}_0}=
\frac{q^{\tilde{E}_0}}{\prod\limits_{j=1}^N (1-q^{2j})} \, ,
\label{A21}
\eeq 
where $q=e^{-1/(k_BT)}$ and $\tilde{E}_0=E_0/a$.

Next, for the purpose of finding out the spectrum and partition function 
of the $BC_N$ type of spin Calogero model with PSRO \eq{A2}, 
let us assume that their exists a projector
$\Lambda^{(m_1,m_2)}$ which would satisfy the relations 
\begin{subequations}
\bea
&&K_{ij}P_{ij} \, \Lambda^{(m_1,m_2)} = \Lambda^{(m_1,m_2)} \,  K_{ij}P_{ij}
= \, -  \, \Lambda^{(m_1,m_2)}  
\label{A7}  \, , \\
&&K_iP_i^{(m_1,m_2)}\Lambda^{(m_1,m_2)} =\Lambda^{(m_1,m_2)} K_i \, 
P_i^{(m_1,m_2)}
=\Lambda^{(m_1,m_2)} \, .
\label{A8}
\eea
\label{A7/A8}
\end{subequations}
Following the procedure of constructing $\Lambda$ \eq{pro4} in 
Sec.~2, we obtain such  $\Lambda^{(m_1,m_2)}$ as 
\beq
\Lambda^{(m_1,m_2)} =
\frac{1}{2^N}\left\{\prod_{j=1}^N\Big(1+ \Pi_j^{(m_1,m_2)}\Big)
\right\}\La_0 \, , 
\label{pro22}
\eeq
where $\Pi_j^{(m_1,m_2)}= K_j P_j^{(m_1,m_2)}$ and $\La_0$ 
is given in \Eq{pro3}. Apart from satisfying the relations \eq{A7/A8}, 
  $\Lambda^{(m_1,m_2)}$ given in \eq{pro22} 
 commutes with the auxiliary operator \eq{auxi}:
\beq
 \left[\Lambda^{(m_1,m_2)}, \mbb{H} \right ] = 0 \, .
\label{comm}
\eeq
With the help of this $\Lambda^{(m_1,m_2)}$,
let us define a set of state vectors depending on both coordinates and 
spins as 
\beq
\psi_{\bf{n}}^{\bf{s}} \equiv 
\psi_{ n_1 ,\ldots, n_i, \ldots, n_j, \ldots , n_N}
^{s_1, \ldots, s_i, \ldots, s_j, \ldots, s_N}=\Lambda^{(m_1,m_2)}
\left (\phi_{\bf{n}} ({\bf x})|{\bf s}\rangle \right) \, ,
\label{A9}
\eeq
where $\phi_{\bf{n}}$ is given in \eq{A10} and 
$\ket{\mbf{s}} \equiv  \ket{s_1,\cdots,s_N}$ is an arbitrary basis element of 
the spin space $\mc{S}$ \eq{basis1}. However, it should be noted that 
$\psi_{\bf{n}}^{\bf{s}}$'s 
defined in \Eq{A9} do not form a set of linearly independent state vectors.
Indeed, by using Eqs.~\eq{A7}, \eq{action1} and \eq{spin4}, 
it is easy to show that
$\psi_{\bf{n}}^{\bf{s}}$'s 
satisfy the antisymmetry condition
\beq
\psi_{ n_1 ,\ldots, n_i, \ldots, n_j, \ldots , n_N}
^{s_1, \ldots, s_i, \ldots, s_j, \ldots, s_N}
=-\psi_{n_1 , \ldots, n_j, \ldots, n_i, \ldots, n_N}^
{s_1, \ldots, s_j, \ldots, s_i, \ldots , s_N} \, .
\label{A15}
\eeq
Furthermore, due to Eqs.~\eq{A8}, \eq{action2} and \eq{diag2},
it follows that  
\beq
\psi_{n_1, \ldots, n_N}^{s_1, \ldots, s_N}=(-1)^{n_i+f(s_i)}~\psi_{n_1,
\ldots, n_N}^{s_1, \ldots, s_N} \, .
\label{A16}
\eeq
The above relation implies that for constructing any nontrivial 
$\psi_{n_1, \ldots, n_N}^{s_1,\ldots, s_N}$, we must take 
$s_i\in\{1, 2, \ldots, m_1\}$ for even values of $n_i$ and  
$s_i\in\{m_1+1, m_1+2, \ldots, m_1+m_2\}$ for odd values of $n_i$.
Using Eqs.~\eq{A15} and \eq{A16} it is easy to check that, 
$\psi_{\bf{n}}^{\bf{s}}$'s 
defined through \Eq{A9} would be nontrivial and linearly independent provided 
the following three conditions are imposed on the corresponding  
$n_i$'s and $s_i$'s.

1) We take an ordered form of $\mbf{n}$, 
which separately arranges its even and odd components 
into two nonincreasing sequences as 
\bea
{\bf n}~\equiv ~(\mbf{n_e}, \mbf {n_o})&=&(\overbrace{2l_1,
\ldots, 2l_1}^{k_1}, \,  \ldots, \,  \overbrace{2l_s, \ldots,
2l_s}^{k_s}, \nn \\ &~& \overbrace{2p_1+1, \ldots, 2p_1+1}^{g_1}, \, 
\ldots, \, \overbrace{2p_t+1,
\ldots, 2p_t+1}^{g_t}) \, ,
\label{A18}
\eea
where  $0\leq s,\, t \leq N$, 
$l_1>l_2>\ldots>l_s\geqslant0$ and $p_1>p_2>\ldots>p_t\geqslant0$.
\vskip 2mm
2) 
The allowed values of $s_i$ corresponding to each $n_i$ are given by 
\beq
s_i\in \left \{
\begin{array}{ll}
 \{1, 2, \ldots, m_1\} \, , & \mbox{for} ~n_i\in \mbf {n_e} \, , \\
\{ m_1+1, m_1+2, \ldots, m_1+m_2\} \, , & \mbox{for}~ n_i\in \mbf {n_o} \, . 
\end{array} 
\right. 
\label{A18b}
\eeq
\vskip 2mm
3) If $n_i=n_j$ and $i<j$, then $s_i>s_j$.
\vskip 2mm
\noindent
We have already discussed how the condition 2) has emerged from \Eq{A16}. 
Due to the condition 2), the numbers of allowed spin components are different
for even and odd values of $n_i$ (except for the particular case where 
$m_1=m_2$, corresponding to even values of $m$). 
Hence, for the sake of convenience, we have taken $\mbf{n}$ in \eq{A18}
 such that its even and odd components are separated
before arranging among themselves. Note that 
 any given $\mbf{n}$ can be brought in the ordered form \eq{A18}
through an appropriate permutation of its 
components. Therefore,  we can impose
the condition 1) as a consequence of \Eq{A15}. 
Finally, the ordering of spin components in condition 3) can also be imposed 
due to \Eq{A15}. However, it should be noted that, 
 the choice \eq{A18} for an ordered form of  
$\mbf{n}$ does not uniquely follow from \Eq{A15}. For example, 
while constructing the basis vectors for the Hilbert space 
of $H^{(m)}$ \eq{ham},  the ordered form \eq{A18} 
of $\mbf{n}$ has been chosen earlier for odd values of $m$,
but  a quite different ordered form of $\mbf{n}$ 
(which arranges all components of $\mbf{n}$ in a nonincreasing sequence, 
without separating them into even and odd parts) 
has been chosen for even values of $m$ \cite{BFGR08}.

All linearly independent
  $\psi_{\bf{n}}^{\bf{s}}$'s \eq{A9},  satisfying  
the above mentioned three conditions,  may now be taken as a set of 
(nonorthonormal) basis vectors for the Hilbert space
of the $BC_N$ type of spin Calogero model with PSRO \eq{A2}. We 
define a partial ordering among these basis vectors 
as: $\psi_{\bf{n}}^{\bf{s}}>\psi_{\bf{n}'}^{ \bf{s}'} \, $, 
~~if~$|\bf{n}|>|\bf{n}'|$.  Using equations~\eq{A7/A8}, \eq{comm}, 
and \eq{tria}, we find that $H^{(m_1,m_2)}$ \eq{A2} acts as an 
upper triangular matrix on these partially ordered basis vectors:
\beq
H^{(m_1,m_2)}\, \psi_{\bf{n}}^{\bf{s}}=E_{\mbf{n}}^{\mbf{s}} \,
 \psi_{\bf{n}}^{\bf{s}} +\sum_{|\bf{m}|<|\bf{n}|}~C_{{\bf mn}} \, 
\psi_{\bf{m}}^{\bf{s}} \, ,
\label{A12}
\eeq
where 
\beq
E_{\mbf{n}}^{\mbf{s}} =a|{\bf{n}}|+E_0 \, .
\label{A13}
\eeq
Due to such triangular form of $H^{(m_1,m_2)}$, all eigenvalues of this   
Hamiltonian are given by \Eq{A13}, where the quantum number $\mbf{n}$ 
satisfies the condition 1) 
and the quantum number $\mbf{s}$ satisfies the conditions 2) and 3). 
Since the right hand side of \Eq{A13} does not depend on the spin 
quantum number $\mbf{s}$, $E_{\mbf{n}}^{\mbf{s}}$'s 
are highly degenerate in general. Using the conditions 2) and 3),
we find out the spin degeneracy 
$d^{\, m_1,m_2}_{\, \bf k, \bf g}$ for the eigenvalue
$E_{\mbf{n}}^{\mbf{s}}$ as 
\beq
d^{\, m_1,m_2}_{\, \bf k, \bf g}
=\prod_{i=1}^sC^{m_1}_{k_i}\prod_{j=1}^tC^{m_2}_{g_j} ~ .
\label{A19}
\eeq
Thus, the degeneracy factors of the energy levels for
 the spectrum of  $H^{(m_1,m_2)}$ \eq{A2}
 explicitly depend on the discrete parameters $m_1$ and $m_2$.  

Since the degree of the monomial $\phi_{\mbf{n}}(\mbf{x})$ \eq{A10} 
with $\mbf{n}$ arranged  
in the form \eq{A18} is given by 
$
|\mbf{n}|= 2\sum_{i=1}^s l_i  k_i+ 2\sum_{j=1}^t p_j g_j + t \, , 
$
the energy eigenvalues \eq{A13} of $H^{(m_1,m_2)}$ can be written as 
\beq
E_{\mbf{n}}^{\mbf{s}} =2a\sum_{i=1}^s l_i  k_i+ 2a\sum_{j=1}^t p_j g_j + at
+E_0 \, .
\label{A13b}
\eeq
Let us denote the numbers of the even and the odd components 
of $\mbf{n}$ by $N_1$ and $N_2$ respectively, 
which can take all possible values ranging from 
$0$ to $N$, and satisfy the condition $N_1 +N_2=N$. 
From \Eq{A18}
it follows that 
\beq
N_1= \sum_{i=1}^s  k_i, ~~~~ N_2 = \sum_{j=1}^t g_j\, . \nn
\eeq
Thus we find that 
${\bf k} \equiv\{k_1, k_2, \ldots, k_s\} \in \mc{P}_{N_1} $ 
and ${\bf g} \equiv \{g_1, g_2, \ldots, g_t\} \in \mc{P}_{N_2} $, 
where $\mc{P}_{N_1}$ and  $\mc{P}_{N_2}$
denote the sets of all ordered partitions of $N_1$ and $N_2$ respectively.
Next, we sum over the Boltzmann weights 
corresponding to all possible $\mbf{n}$  in the ordered form \eq{A18},
by using the corresponding energy eigenvalues  \eq{A13b} and their  
degeneracy factors  \eq{A19}.
Thus we obtain the canonical partition function for the
$BC_N$ type of spin Calogero model with PSRO 
\eq{A2} as
\beq
Z_N^{(m_1,m_2)}(aT) =q^{\tilde{E}_0} \! \! \! \!
\sum_{\stackrel{N_1,N_2}{(N_1+N_2=N)}}
 \sum_{{\bf k} \in \mc{P}_{N_1}, \, {\bf g} \in \mc{P}_{N_2}} \! \!
d^{\, m_1,m_2}_{\, \bf k, \bf g} \! \!
\sum_{l_1>\dots>l_s \geq
0} ~\sum_{p_1>\dots>p_t \geq 0}    q^{2 \sum\limits_{i=1}^s l_i  k_i+ 
2\sum\limits_{j=1}^t p_j g_j +t} \, , \nn 
\eeq
where $q=e^{-1/(k_BT)}$  and $\tilde{E}_0 = E_0/a$. 
Summing over $l_i$'s and $p_j$'s through 
appropriate change of variables, as done in Ref~\cite{BFGR08}
while calculating the partition function of $H^{(m)}$ \eq{ham}
for odd values of $m$,   we get a simplified expression
for the above partition function as 
\beq
Z_N^{(m_1,m_2)}(aT) =q^{\tilde{E}_0}
\sum_{\stackrel{N_1,N_2}{(N_1+N_2=N)}}
 \sum_{{\bf k} \in \mc{P}_{N_1}, \, {\bf g} \in \mc{P}_{N_2}}
d^{\, m_1,m_2}_{\, \bf k, \bf g} \,
q^{-(N+\kappa_s)} \prod_{i=1}^s
\frac{q^{2\kappa_i}}{1-q^{2\kappa_i}} 
\prod_{j=1}^t\frac{q^{2\zeta_j}}{1-q^{2\zeta_j}} \, ,
\label{A20}
\eeq
where $\kappa_i =\sum_{l=1}^i k_l$ and $\zeta_j =\sum_{l=1}^j g_l$ 
denote the partial sums corresponding to the partitions 
${\bf k} \in \mc{P}_{N_1}$ and 
${\bf g}\in \mc{P}_{N_2}$ respectively. 
Using Eqs.~\eq{A6}, \eq{A21} and \eq{A20}, 
 we finally obtain an expression for the partition function of
the $BC_N$ type of PF spin chain with PSRO \eq{A4} as 
\beq
\mc{Z}_N^{(m_1, m_2)}(T)=
\prod_{l=1}^N (1-q^{2l})
\sum_{\stackrel{N_1,N_2}{(N_1+N_2=N)}}
 \sum_{{\bf k} \in \mc{P}_{N_1}, \, {\bf g} \in \mc{P}_{N_2}}
d^{\, m_1,m_2}_{\, \bf k, \bf g} \,
q^{-(N+\kappa_s)} \prod_{i=1}^s
\frac{q^{2\kappa_i}}{1-q^{2\kappa_i}} 
\prod_{j=1}^t\frac{q^{2\zeta_j}}{1-q^{2\zeta_j}} \, .
\label{A22}
\eeq 
However, from the above equation it is not clear whether 
$\mc{Z}_N^{(m_1, m_2)}(q)$ 
can be expressed as a polynomial function of $q$, 
which is expected for the case of any 
finite system with integer energies. In the following, we shall try to express
$\mc{Z}_N^{(m_1, m_2)}(q)$ as a polynomial of $q$ by using
the $q$-binomial coefficients.
To this end, we define complementary sets of the two sets  
$\{\kappa_1, \kappa_2, \ldots, \kappa_s\}$ and 
$\{\zeta_1, \zeta_2, \ldots, \zeta_t\}$
as $\{1, 2, \ldots, N_1-1, N_1\}-\{\kappa_1, \kappa_2,
\ldots, \kappa_s\}\equiv \{\kappa_1', \kappa_2',
\ldots, \kappa'_{N_1-s}\}$  and $\{1, 2, \ldots, N_2-1, N_2\}-
\{\zeta_1, \zeta_2, \ldots, \zeta_t\}
\equiv \{\zeta_1', \zeta_2', \ldots, \zeta'_{N_2-t}\}$, respectively.
Using the elements belonging to these complementary sets,
one can write  
\beq
\prod_{i=1}^s\frac{1}{1-q^{2\kappa_i}}=\frac{\prod\limits_{i=1}^{N_1-s}(1-
q^{2\kappa'_i})}{\prod\limits_{ i=1}^{N_1}(1-q^{2i})} ~ , ~~~~
\prod_{j=1}^t\frac{1}{1-q^{2\zeta_j}}=
\frac{\prod\limits_{j=1}^{N_2-t}(1-q^{2\zeta'_j})}{
\prod\limits_{ j=1}^{N_2}(1-q^{2j})} \, .
\label{A24}
\eeq
Substituting (\ref{A24})  to (\ref{A22}),  we get
\bea
\mc{Z}_N^{(m_1, m_2)}(T) &=&
\sum_{\stackrel{N_1,N_2}{(N_1+N_2=N)}}
 \sum_{{\bf k} \in \mc{P}_{N_1}, \, {\bf g} \in \mc{P}_{N_2}}
d^{\, m_1,m_2}_{\, \bf k, \bf g} \,
q^{-(N+\kappa_s)+2\sum\limits_{j=1}^s \kappa_j+2\sum\limits_{j=1}^t
\zeta_j} \nn \\
&~&
~~~~~~~~~~~\times \prod\limits_{i=1}^{N_1-s}(1-q^{2\kappa'_i}) 
\prod\limits_{j=1}^{N_2-t}(1-q^{2\zeta'_j}) 
 \frac{\prod\limits_{l=1}^N
(1-q^{2l})}{\prod\limits_{ i=1}^{N_1}(1-q^{2i})\prod\limits_{
j=1}^{N_2}(1-q^{2j})}  \, . \nn
\eea
Since $\kappa_s =N_1$ and $\zeta_t =N_2$, the above equation can also 
be expressed as 
\beq
\mc{Z}_N^{(m_1, m_2)}(T)  = \hskip -.4 cm  
\sum_{\stackrel{N_1,N_2}{(N_1+N_2=N)}}
 \sum_{{\bf k} \in \mc{P}_{N_1}, \, {\bf g} \in \mc{P}_{N_2}}
\hskip -.4 cm 
d^{\, m_1,m_2}_{\, \bf k, \bf g} \,
q^{N_2+2\sum\limits_{i=1}^{s-1} \kappa_i+2\sum\limits_{j=1}^{t-1}
\zeta_j}
\prod\limits_{i=1}^{N_1-s}(1-q^{2\kappa'_i})
\prod\limits_{j=1}^{N_2-t}(1-q^{2\zeta'_j})\qbinom N{N_1}{q^2} \, , 
\label{A25}
\eeq
where $\qbinom N{N_1}{q^2}$ denotes a $q$-binomial coefficient defined by 
\beq
\qbinom N{N_1}{q^2}= \frac{\prod\limits_{l=1}^N
(1-q^{2l})}{\prod\limits_{ i=1}^{N_1}(1-q^{2i})\prod\limits_{
j=1}^{N-N_1}(1-q^{2j})} \, .  \nn
\eeq
It is well known that a $q$-binomial coefficient like  $\qbinom N{N_1}{q^2}$
can be written as an even polynomial of degree $2N_1(N-N_1)$ in $q$ 
\cite{Ci79}. 
Hence, the partition function \eq{A25} of the 
$BC_N$ type of PF spin chain with PSRO \eq{A4} is finally
expressed as a polynomial in $q$. Since the  partition function \eq{A25}
does not depend on the parameter $\beta$ which is present 
in the Hamiltonian \eq{A4}, it is evident that 
the energy levels of this Hamiltonian 
do not change with the variation of $\beta$. 

Let us now compare the partition function \eq{A25} with the previously 
obtained partition function \cite{BFGR08} of the spin chain \eq{A4b}.
As expected, in the special given by  
$m_1=m_2+\ep$ for odd values of $m$,
\eq{A25} reproduces the partition 
function  of the spin chain \eq{A4b}. 
However, in the special case given by  
$m_1=m_2$ for even values of $m$,
\eq{A25} yields an equivalent but apparently different looking expression
for the partition function of the spin chain \eq{A4b}.
This happens because
the ordering of $\mbf{n}$, which was chosen earlier 
while computing the partition function  of the spin chain \eq{A4b}, 
is same as \eq{A18} for odd values of $m$, 
but different from \eq{A18} for even values of $m$.
It may also be noted that, for even values of $m$, 
the partition function of the $BC_N$ type of PF spin chain \eq{A4b}
can be related in a very simple way  
to the partition function of the $A_{N-1}$ type of PF spin chain \eq{a1}
with $m/2$ number of internal degrees of freedom \cite{BFGR08}. 
However, no such simple relation is known to exist between the 
partition functions of the $BC_N$ and $A_{N-1}$ types of PF spin chains
for odd values of $m$. In the following section, we shall establish a 
novel relation between the partition functions of the 
 $BC_N$ type of PF spin chains with PSRO 
and $A_{N-1}$ type of PF spin chain, 
which would remain uniformly valid for all possible choice of $m_1$ and $m_2$ 
corresponding to  both even and odd values of $m$.

\bigskip

\noi \section{Relation with the partition function of $A_{N-1}$ type 
PF spin chain}
\renewcommand{\theequation}{4.{\arabic{equation}}}
\setcounter{equation}{0}
\medskip

For the purpose of making a connection between the partition function \eq{A25}
of the $BC_N$ type of PF spin chain with PSRO
and that of the $A_{N-1}$ type of PF spin chain,
at first we observe that the spin degeneracy factor
$d^{\, m_1,m_2}_{\, \bf k, \bf g}$ \eq{A19} may be written as 
\beq
d^{\, m_1,m_2}_{\, \bf k, \bf g}
=d_{m_1}({\bf k})d_{m_2}({\bf g}) \, ,
\label{A19f}
\eeq
where 
\beq
d_{m_1}({\bf k})= \prod_{i=1}^sC^{m_1}_{k_i} , ~~~~
d_{m_2}({\bf g})= \prod_{j=1}^tC^{m_2}_{g_j} \, . \nn
\eeq
Substituting $d^{\, m_1,m_2}_{\, \bf k, \bf g}$ in \Eq{A19f} to \Eq{A25},
we obtain
\bea
&&\mc{Z}_N^{(m_1, m_2)}(T)
= \sum_{\stackrel{N_1,N_2}{(N_1+N_2=N)}}
q^{N_2} \qbinom N{N_1}{q^2} \left(  \sum_{ {\bf k} \in 
\mc{P}_{N_1}} d_{m_1}({\bf
k})q^{2 \sum_{j=1}^{s-1} \kappa_j} \prod_{j=1}^{N_1-s}(1- q^{2 \kappa_j'}) \right)\nn \\
&&~~~~~~~~~~~~~~~~~~~~~~~~~~~~~~~~~~~~~~~~
\times \left(  \sum_{ {\bf g} \in \mc{P}_{N_2}} \,   d_{m_2}({\bf
g}) q^{2 \sum_{j=1}^{t-1} \zeta_j} \prod_{j=1}^{N_2-t}(1- q^{2 \zeta_j'}) \right) \, .\label{ANBN1}
\eea
In this context it may be noted that,  there exists several different but 
equivalent expressions for the partition function of the 
 $A_{N-1}$ type of PF spin chain \eq{a1} in the literature
 \cite{Po94,Hi95npb, BFGR08epl, BBHS07}. 
For our present purpose,
we shall use the following expression \cite{BFGR08epl, BBHS07}
for the partition function of the 
 $A_{N-1}$ type of PF spin chain \eq{a1} with $m$ internal degrees of freedom:  
\beq
\mathcal{Z}^{A,m}_{N}(T)= \sum_{ {\bf f} \in 
\mc{P}_{N}} d_{m}({\bf f})q^{ \sum_{j=1}^{r-1} \mc{F}_j} 
\prod_{j=1}^{N-r}(1- q^{\mc{F}_j'}) \, .
 \label{A27}
\eeq
where $\mbf{f} \equiv \{f_1, f_2 \cdots f_r\}$, 
$d_{m}({\bf f})= \prod_{i=1}^rC^{m}_{f_i}$, 
the 
partial sums are given by $\mc{F}_j = \sum_{i=1}^j f_i$, and 
 the complementary partial sums are defined as
$\{ \mc{F}_1', \mc{F}_1' , \cdots ,  \mc{F}_{N-r}' \}
\equiv \{1, 2, \cdots, N\}-\{\mc{F}_1, \mc{F}_2,
\cdots, \mc{F}_r\}$. Let us now 
multiply $ \mc{H}_\mr{PF}^{(m)}$ in \eq{a1} by a factor of two and define  
a scaled Hamiltonian for the
$A_{N-1}$ type of PF spin chain as 
 \beq
\wt{\mc{H}}_\mr{PF}^{(m)} \equiv 
2\mc{H}_\mr{PF}^{(m)}=\sum_{i\neq j}
\frac{1+P_{ij}}{(\rho_i-\rho_j)^2}  \, . 
\label{scal}
\eeq
Since all energy levels of $\wt{\mc{H}}_\mr{PF}^{(m)}$ 
are related to those of  $\mc{H}_\mr{PF}^{(m)}$ 
by a scale factor of two, the partition 
function of $\wt{\mc{H}}_\mr{PF}^{(m)}$ 
(which is denoted by $\wt{\mc{Z}}^{A,m}_{N}(T)$) can be obtained 
from the r.h.s. of \Eq{A27} by simply substituting  
$q^2$ to the place of $q$:  
\beq
\wt{\mc{Z}}^{A,m}_{N}(T)= \sum_{ {\bf f} \in 
\mc{P}_{N}} d_{m}({\bf f})q^{2 \sum_{j=1}^{r-1} \mc{F}_j} 
\prod_{j=1}^{N-r}(1- q^{2\mc{F}_j'}) \, .
 \label{A27b}
\eeq
Using \eq{A27b}, we finally express 
$\mc{Z}_N^{(m_1, m_2)}(T)$ in \eq{ANBN1} as
\beq
\mc{Z}_N^{(m_1, m_2)}(T)
= \sum_{N_1=0}^N q^{N-N_1} \qbinom N{N_1}{q^2} 
\wt{\mathcal{Z}}^{A,m_1}_{N_1}(T)\,
\wt{\mathcal{Z}}^{A,m_2}_{N-N_1}(T) \, .
\label{ANBN2}
\eeq
Thus we obtain a remarkable 
relation between the partition function of the 
 $BC_N$ type of PF spin chain with PSRO 
and the partition functions of several $A_{N-1}$ type of PF spin chains,  
which can be applied for all possible values of $m_1$ and $m_2$.

However, it should be observed that, even in the special cases like 
$m_1=m_2$ for even values of $m$, our relation \eq{ANBN2}
does not coincide with the previously derived relation \cite{BFGR08} between the 
partition functions of the $BC_N$  and $A_{N-1}$ types of PF spin chains. 
To shed some light on this matter through a particular example, 
let us choose the simplest case given by  $m_1=m_2=1$ for $m=2$. 
Since $d_{1}({\bf f})=1$ for $\mbf{f}= \{1,1, \cdots, 1\} \in \mc{P}_{N} $, 
and  $d_{1}({\bf f})=0$ for any other $\mbf{f} $ within the set $\mc{P}_{N}$, 
from \Eq{A27b} it follows that $\wt{\mc{Z}}^{A,1}_{N}(T)=q^{N(N-1)}$.  
Hence, by putting $m_1=m_2=1$ in \Eq{ANBN2}, we find that  
\beq
\mc{Z}_N^{(1, 1)}(T)
= \sum_{N_1=0}^N q^{(N-N_1)^2+N_1(N_1-1)}\qbinom N{N_1}{q^2} .
\label{l6}
\eeq
As has been mentioned earlier, in the particular case given by $m_1=m_2=1$, 
the $BC_N$ type of PF spin chain with PSRO \eq{A4} reduces to the 
$BC_N$ type of PF spin chain \eq{A4b} with $m=2$. For this case, 
the previously derived relation  between the 
partition functions of the $BC_N$ and $A_{N-1}$ types of PF spin chains
yields \cite{BFGR08} 
\beq
\mc{Z}_N^{(1, 1)}(T)
=q^{\frac{N(N-1)}{2}} \prod_{i=1}^N (1+q^i) .
\label{l7}
\eeq
Comparing the r.h.s. of Eqs.~(\ref{l6}) and (\ref{l7}), 
we obtain an interesting identity of the form 
\beq
\prod_{i=1}^N (1+q^i) = \sum_{l=0}^N q^{\frac{(N-2l)(N-2l+1)}{2}}\qbinom N{l}
{q^2}. \nn
\eeq

Let us now consider another particular case  
given by $m_1=m$, $m_2=0$, for which the $BC_N$ type of PF spin chain
with PSRO \eq{A4} 
reduces to the SA type generalization \eq{a2} of the PF spin chain. Due to 
\Eq{A18b}, it is evident that there exists no odd sector of $\mbf{n}$
in \eq{A18}, i.e., $N_2=0$ in this case.
 Therefore, the summation variable $N_1$ 
 can only take the value $N$ (instead of its usual range from $0$ to $N$)
in the r.h.s. of Eqs.~\eq{ANBN1} and \eq{ANBN2}. Consequently, in this 
special case, \Eq{ANBN2} yields  
\beq
\mc{Z}_N^{(m, 0)}(T)
=\wt{\mathcal{Z}}^{A,m}_{N}(T)\, .
\label{A35}
\eeq
The above equality between two partition functions implies that 
the spectrum of the SA type generalization of the PF spin chain \eq{a2}
with arbitrary value of the parameter $\beta$
is exactly same with that of the $A_{N-1}$ type of PF spin chain \eq{scal}.
This result is quite surprising, since the form of the two Hamiltonians 
given in \eq{a2} and \eq{scal} apparently differ from each other. 
Indeed, only in the simplest case of $N=2$, 
we are able to analytically show that the two Hamiltonians 
given in \eq{a2} and \eq{scal} coincide with each other for any value of 
$\beta$. On the other hand, by ordering the zero points of the Hermite
polynomial $H_N$ and the generalized Laguerre polynomial $L_N^{\beta -1}$
on the real line as  
$\rho_1 > \rho_2> \cdots > \rho_N$ and $y_1>y_2>\cdots >y_N$ respectively, 
one can numerically verify that the following inequalities hold
for finite values of $\beta$ and for some $N\geq 3$: 
\beq
\frac{y_i+y_j}{(y_i-y_j)^2} \neq \frac{1}{(\rho_i -\rho_j)^2} \, , 
\eeq
where $1\leq i<j\leq N$. Even though the above inequalities hold 
for finite values of $\beta$, things become more interesting
 in the limit of $\beta $ tending to infinity.  
In fact, we numerically find that the asymptotic relations 
given by   
\beq
 \underset{\beta \rightarrow \infty    }{\mr{Lim}}   \,
 \frac{y_i+y_j}{(y_i-y_j)^2} = \frac{1}{(\rho_i -\rho_j)^2} \, , 
 \label{A38}
\eeq
where $1\leq i<j\leq N$,  hold for $N=3$ and $N=4$ cases.  Being encouraged 
by such numerical evidence, we  
 conjecture that the asymptotic relations given in  
Eq. (\ref{A38}) hold for arbitrary values of $N$. 
This conjecture clearly implies that 
\beq
\wt{\mc{H}}_\mr{PF}^{(m)} 
=\underset{\beta \rightarrow \infty    }{\mr{Lim}}  \, 
{\mc{H}}^{(m,0)} \, ,
\label{A39}
\eeq
i.e., the scaled Hamiltonian  \eq{scal} of the 
$A_{N-1}$ type of PF spin chain may be seen as a particular limit 
of the Hamiltonian \eq{a2}
corresponding to the SA type generalization of the PF spin chain.
Moreover, since the spectrum of ${\mc{H}}^{(m,0)}$ does not depend 
on the value of $\beta$, this Hamiltonian may be interpreted as 
an isospectral deformation of $\wt{\mc{H}}_\mr{PF}^{(m)}$.


It is well known that the partition functions of the $A_{N-1}$ type of
ferromagnetic and anti-ferromagnetic PF spin chains satisfy a duality
relation \cite{Po94,BUW99,HB00}. This type of duality relation 
has also been established for the case of $BC_N$ type of 
anti-ferromagnetic PF spin chain \eq{A4b} and its ferromagnetic
counterpart \cite{BFGR08}. Since the partition functions 
of the $BC_N$ type of PF spin chains with PSRO can be expressed through    
the partition functions of the $A_{N-1}$ type of PF spin chains, 
it is expected that the partition functions of the former type of
ferromagnetic and anti-ferromagnetic spin chains 
would also satisfy a duality relation. 
For the purpose of finding out such duality relation,  
we define the ferromagnetic counterpart corresponding to the 
$BC_N$ type of anti-ferromagnetic PF spin chain with PSRO (\ref{A4}) as
\beq
\wh{\mathcal{H}}^{(m_2,m_1)}=\sum_{i\neq j}
\left[\frac{1-P_{ij}}{(\xi_i-\xi_j)^2} +
\frac{1-\widetilde{P}_{ij}^{(m_2,m_1)}}{(\xi_i+\xi_j)^2}\right]
+\beta \sum_{i=1}^{N}\frac{1-P_i^{(m_2,m_1)}}{\xi_i^2}.
\label{A28}
\eeq
Next, by using  \Eq{trace1}, we  find that trace of the  
operators $P_i^{(m_2,m_1)}$ and  $- P_i^{(m_1,m_2)}$ coincide 
with each other. Since the eigenvalues of both 
$P_i^{(m_2,m_1)}$ and  $- P_i^{(m_1,m_2)}$ can only be $\pm 1$, 
these two operators with exactly same eigenvalues 
must be related through a similarity  transformation. Hence,
there exists a symmetric operator $M$ such that
\beq
M P_i^{(m_2,m_1)}M^{-1}=-P_i^{(m_1,m_2)}  , ~~M P_{ij}M^{-1}=P_{ij}  , 
~~  M \wt{P}_{ij}^{(m_1,m_2)}M^{-1}=\wt{P}_{ij}^{(m_2,m_1)} \, . 
\label{A29}
\eeq
Using Eqs.~\eq{A28} and \eq{A29}, we get
\beq
M\wh{\mathcal{H}}^{(m_2,m_1)}M^{-1}=\sum_{i\neq j}
\left[\frac{1-P_{ij}}{(\xi_i-\xi_j)^2} +
\frac{1-\widetilde{P}_{ij}^{(m_1,m_2)}}{(\xi_i+\xi_j)^2}\right]
+\beta \sum_{i=1}^{N}\frac{1+P_i^{(m_1,m_2)}}{\xi_i^2}.
\label{A30}
\eeq
Adding up the expressions in Eqs.~(\ref{A4}) and (\ref{A30}), we obtain
\beq
\mathcal{H}^{(m_1,m_2)}+M\wh{\mathcal{H}}^{(m_2,m_1)}M^{-1}
=2 \sum_{i\neq j} (h_{ij} + \tilde{h}_{ij}) + 
\beta \sum_{i=1}^N \frac{1}{y_i} \, , 
\label{sumeq}
\eeq
where $h_{ij}=\frac{1}{(\xi_i-\xi_j)^2}$ and 
$\widetilde{h}_{ij}=\frac{1}{(\xi_i+\xi_j)^2}$. 
Using the relation \eq{glag} satisfied by the zero points 
of the generalized Laguerre polynomial, it can be shown that 
\cite{Ah78,AM83,BFGR08}
 \beq
 \sum_{i\neq j}(h_{ij} + \tilde{h}_{ij}) = \frac{N(N-1)}{2} \, , ~~
\sum_{i=1}^N \frac{1}{y_i}= \frac{N}{\beta} \, .
\label{iden}
\eeq
Consequently, \Eq{sumeq} can be written as 
\beq
\mc{H}^{(m_1,m_2)}+M\wh{\mc{H}}^{(m_2,m_1)}M^{-1}=N^2.
\label{A31}
\eeq
Since $\wh{\mc{H}}^{(m_2,m_1)}$ and 
$M\wh{\mc{H}}^{(m_2,m_1)}M^{-1}$ are
isospectral Hamiltonians, from the above equation it follows that 
\beq
\wh{\mc{E}}_j = N^2 - \mc{E}_j \, ,
\label{A31b} 
\eeq
where $\mc{E}_j$ and $\wh{\mc{E}}_j$
denote the eigenvalues of  $\mc{H}^{(m_1,m_2)}$
and $\wh{\mc{H}}^{(m_2,m_1)}$ respectively. 
Due to \Eq{A31b}, there exists a one-to-one correspondence between  
the eigenvalues of  $\mc{H}^{(m_1,m_2)}$ and 
those of $\wh{\mc{H}}^{(m_2,m_1)}$. 
Hence, one can easily derive a 
duality relation between the partition functions 
 of the anti-ferromagnetic spin chain \eq{A28} and 
that of the ferromagnetic spin chain \eq{A4}  as 
\beq
\wh{\mc{Z}}_{N}^{(m_2,m_1)}(T)=q^{N^2} \, 
\mc{Z}_{N}^{(m_1,m_2)}(T)|_{q\to q^{-1}} \, ,
\label{A32}
\eeq
where $\wh{\mc{Z}}_{N}^{(m_2,m_1)}(T)$ denotes the partition function 
of the anti-ferromagnetic spin chain. Since
$\mc{Z}_{N}^{(m_1,m_2)}(T)|_{q\to q^{-1}}$  may be obtained from the 
r.h.s. of \Eq{A25} after replacing $q$ by $q^{-1}$, the duality relation
\eq{A32} can be used to derive an expression for the partition function 
of the anti-ferromagnetic spin chain \eq{A28}. 

\bigskip

\noi \section{Ground state and highest state energies 
for spin chains with PSRO}
\renewcommand{\theequation}{5.{\arabic{equation}}}
\setcounter{equation}{0}
\medskip
In the present section, at first our aim is to  
calculate the ground state energy 
$\mathcal{E}_{min}$ of the $BC_N$ type of anti-ferromagnetic 
PF spin chain with PSRO \eq{A4} by using the freezing trick. 
To this end, we consider \Eq{deco} which implies that 
\beq
\mathcal{E}_{min}=\lim_{a \rightarrow \infty}\frac{1}{a}(E_{min}-E^{sc}_{min}),
\label{minen}
\eeq
where $E^{sc}_{min}$ and $E_{min}$ represent the ground state energies 
of the $BC_N$ type of scalar Calogero model \eq{A5a} 
 and spin Calogero model with PSRO \eq{A2}, respectively. It has been  
mentioned earlier that the eigenvalues of the $BC_N$ type of 
scalar Calogero model are given by 
\Eq{Ep}, where $n_i$'s are even integers satisfying the relation  
$n_1 \geq n_2 \geq \cdots \geq n_N \geq 0$. Hence, by 
choosing all $n_i$ as zero,  one finds that $E^{sc}_{min}=E_0$. 
Due to \Eq{A13}, we can express the  ground state 
energy of spin Calogero model with PSRO
as $E_{min}= a|{\bf{n}}|_{min}+E_0$, 
where $|{\bf{n}}|_{min}$
represents the minimum value of $|{\bf{n}}|$
 for all possible choice of the multi-index ${\bf n}$
compatible with the conditions $1)-3)$ of section 3.  
Substituting these expressions of $E_{min}$ and $E^{sc}_{min}$ in 
 \Eq{minen}, we obtain
\beq
\mathcal{E}_{min}=|{\bf{n}}|_{min} \, .
\label{minen2}
\eeq

For the purpose of calculating $|{\bf{n}}|_{min}$, it is convenient
to consider two different ranges of the number $l$ defined by 
$l=N\, \text{mod}~ m$.
Evidently, $N$ can be expressed through $l$ as
\beq
N= k m +l \, ,
\label{ldef2}
\eeq 
where $k$ is a nonnegative integer. For the case $0 \leq l<m_1$, 
let us construct
a multi-index $\mbf{n}$ by combining the following 
even and odd components according to \eq{A18}:
\bea
&&\mbf{n_e}= \overbrace{2k,\ldots,2k}^{l}, \, 
\overbrace{2(k-1),\ldots,2(k-1)}^{m_1}, ~ \ldots  ~ , \, 
\overbrace{2,\ldots,2}^{m_1}, \, 
\overbrace{0,\ldots,0}^{m_1} \, ,~ \nn \\
&&\mbf {n_o}=
\overbrace{2k-1,\ldots,2k-1}^{m_2}, ~\ldots ~, \, 
\overbrace{3,\ldots,3}^{m_2}, \, 
\overbrace{1,\ldots,1}^{m_2} \, . \nn
\eea
Applying the conditions 2) and 3) of section 3, 
it is easy to check that such $\mbf{n}$ yields 
 $|{\bf{n}}|_{min}$ with value given by 
\beq
|{\bf{n}}|_{min}
=k\big\{ (k-1)m+2l+m_2 \big\} \, .
\label{energy1}
\eeq
Using Eqs.~\eq{minen2}, \eq{energy1} and \eq{ldef2}, we express
the ground state energy of the anti-ferromagnetic 
spin chain with PSRO \eq{A4} as  
\beq
\mathcal{E}_{min}=\frac{1}{m}(N-l)(N+l-m_1), ~~~ \text{where} ~~0\leq l < m_1.
\label{energy2}
\eeq
Subsequently, for the case $m_1 \leq l<m$, we construct
a multi-index $\mbf{n}$ by combining the following 
even and odd components according to \eq{A18}:
\bea
&&\mbf{n_e}= \overbrace{2k,\ldots,2k}^{m_1},~ \ldots ~, \, 
\overbrace{2,\ldots,2}^{m_1} \, , \, 
\overbrace{0,\ldots,0}^{m_1} \, ,~ \nn \\
&&\mbf {n_o}=
\overbrace{2k+1,\ldots,2k+1}^{l-m_1}, \, 
\overbrace{2k-1,\ldots,2k-1}^{m_2}, ~\ldots , \,
\overbrace{3,\ldots,3}^{m_2}, \, 
\overbrace{1,\ldots,1}^{m_2} \, . \nn
\eea
Again, applying the conditions 2) and 3) of section 3, 
we find that such $\mbf{n}$ yields 
 $|{\bf{n}}|_{min}$ with value given by 
\beq
|{\bf{n}}|_{min}
=k\big\{ (k-1)m+2l+m_2 \big\} + (l-m_1)\, .
\label{energy3}
\eeq
Using Eqs.~\eq{minen2}, \eq{energy3} and \eq{ldef2}, we obtain 
the ground state energy of the anti-ferromagnetic 
spin chain \eq{A4} as 
\beq
\mathcal{E}_{min}=\frac{1}{m}(N-l)(N+l-m_1)+(l-m_1),~~~ \text{where}~~ 
 m_1 \leq l<m  .
\label{energy4}
\eeq
It is easy to check that in the special case given by   
$m_1=m_2$ ($m_1=m_2+\ep$) for even (odd) values of $m$,
Eqs.~\eq{energy2} and \eq{energy4} reproduce the ground state energy
obtained in Ref.~\cite{BFGR08}
for the spin chain \eq{A4b}. 
Next, let us consider another
special case given by  $m_1 =m, ~m_2=0$, 
for which the spin chain with PSRO \eq{A4} 
reduces to the SA type generalization \eq{a2} of the PF spin chain.
It is evident that \Eq{energy4} is not
relevant for this case. Hence, by using \Eq{energy2}, we obtain the 
ground state energy of the spin chain \eq{a2} as 
\bea
\mathcal{E}_{min}=\frac{(N-l)(N+l-m)}{m},~ 
\text{where}~ l\equiv N \, \text{mod} \, m \, ,
\label{fer}
\eea
which, as expected, is exactly double of the 
ground state energy associated with the $A_{N-1}$ type of
anti-ferromagnetic PF spin chain \eq{a1} \cite{BFGR08epl}. 

Next, we want to find out the highest energy level
$\mathcal{E}_{max}$ for the $BC_N$ type of anti-ferromagnetic 
PF spin chain with PSRO \eq{A4}. 
Since $P_{ij}^2=\big(\tilde{P}_{ij}^{(m_1,m_2)}\big)^2
=\big(P_i^{(m_1,m_2)}\big)^2= \one $, 
each of these operators can have the eigenvalues $\pm1$.
If there exists a simultaneous eigenstate of these operators
 such that the eigenvalues of 
$P_{ij}$, $\tilde{P}_{ij}^{(m_1,m_2)}$ and 
$P_i^{(m_1,m_2)}$ are given by $+1$, $+1$ and $-1$ respectively, 
 then that eigenstate would evidently yield the highest energy eigenvalue
for the spin chain \eq{A4}. 
For the case of an arbitrary value of $m_1$ and $m_2 >0$, 
we can easily construct such an eigenstate as
$\ket{s,s, \cdots , s}$, where $s>m_1$.  
Hence, by using \Eq{A4}, we get  
\beq
\mathcal{E}_{max}=2 \sum_{i\neq j} (h_{ij} + \tilde{h}_{ij}) + 
\beta \sum_{i=1}^N \frac{1}{y_i} \, . 
\label{sumeq1}
\eeq
Using the identities given in \Eq{iden}, we obtain
the highest energy eigenvalue for the spin chain \eq{A4}  
in $m_2 > 0$ case as 
\beq
\mathcal{E}_{max} =N^2.
\label{nzq}
\eeq
Next, let us consider the case given by  
$m_1 =m, ~m_2=0$. In this case, the operator
$P_i^{(m_1,m_2)}$ is not allowed to take the eigenvalue $-1$. 
Hence, if we consider a spin state like 
 $\ket{s,s, \cdots , s}$, with $0\leq s \leq m$,  
the eigenvalues of all of the operators  
$P_{ij}$, $\tilde{P}_{ij}^{(m_1,m_2)}$ and 
$P_i^{(m_1,m_2)}$  would be given by $+1$. 
Consequently, by using \Eq{A4} and the first identity given in \Eq{iden}, 
we obtain the highest energy eigenvalue for the spin chain \eq{A4} in 
the case $m_2 = 0$ as 
\beq
\mathcal{E}_{max}=N(N-1).
\label{m2}
\eeq
Again, it is interesting to note that this 
$\mathcal{E}_{max}$ is exactly double of the 
highest energy eigenvalue associated with the $A_{N-1}$ type of
anti-ferromagnetic PF spin chain \eq{a1} \cite{BFGR08epl}.

It is well known that the spectrum of the  
$A_{N-1}$ type of PF spin chain \eq{a1} is equispaced within
its lowest and highest energy levels. This result 
follows from the fact that corresponding partition function  
$\mathcal{Z}^{A,m}_{N}(T)$ \eq{A27} can be expressed as a polynomial in
$q$ with degree $N(N-1)/2$, where all consecutive powers of $q$
(within the allowed range) appear with 
positive integer coefficients \cite{Hi95npb}. 
In this context, it is interesting to ask whether the spectrum of the   
$BC_N$ type of PF spin chain with PSRO \eq{A4} is also equispaced.
To answer this question, let us first consider the 
special case given by  $m_1 =m, ~m_2=0$.
Using \Eq{A35} for this special case, we find that the 
corresponding partition function can be expressed as a polynomial in
$q$ with degree $N(N-1)$, where all consecutive $\it{even}$ powers of $q$ 
appear with positive integer coefficients.
Hence the spectrum of the spin chain \eq{A4} is equispaced 
in the above mentioned special case. Next, for the purpose
of finding out the nature of spectrum in the case $m_2 \neq 0$,     
we examine all terms appearing in the
 corresponding partition function \eq{ANBN2}.
It may be noted that, $\qbinom N{N_1}{q^2}$,
$\tilde{\mathcal{Z}}^{A,m_1}_{N_1}(T)$ and
$\tilde{\mathcal{Z}}^{A,m_2}_{N-N_1}(T)$ can be expressed as polynomials of $q$,
where all consecutive $\it{even}$ powers of $q$ (within appropriate ranges) 
appear with positive integer coefficients.
However, the first factor of the summand in the r.h.s. of \Eq{ANBN2}
is a monomial in $q$ which, as the
summation runs, takes all odd and even powers within the range $0$ to $N$. 
Consequently, $\mc{Z}_N^{(m_1, m_2)}(T)$ in \Eq{ANBN2}
can be expressed as a polynomial in $q$, where all 
possible consecutive  powers of $q$ appear with positive integer coefficients.
Hence, the spectrum of the spin chain \eq{A4} is equispaced 
also for the case $m_2 \neq 0$. However, in this case, the spacing between two 
consecutive levels reduces by a factor of half in comparison with 
that of the $m_2=0$ case. 

\bigskip

\noi \section{Spectral properties of the spin chains with PSRO}
\renewcommand{\theequation}{6.{\arabic{equation}}}
\setcounter{equation}{0}
\medskip

In this section we shall study a few spectral  
properties of the $BC_N$ type of PF spin chain with PSRO \eq{A4},
like its energy level density and   
nearest neighbour spacing distribution, for the case of finite 
but sufficiently large number of lattice sites. 
It was observed earlier \cite{BFGR08epl,BFGR08} that, 
for sufficiently large number of lattice sites, 
the energy level densities of both $A_{N-1}$ type of PF spin chain \eq{a1}
and $BC_N$ type of PF spin chain \eq{A4b} tend to follow   
the Gaussian distribution with high
degree of accuracy. 
An analytical proof for the Gaussian behaviour of the level
density distributions at $N\to \infty$ limit was given  
for the case of $A_{N-1}$ type of spin chains and related 
one-dimensional vertex models \cite{EFG10, BB12}.
It was also found that, in contrast to the case of 
some other integrable systems \cite{PZBMM93, AMV02}, 
the spacings between consecutive levels in the spectra 
of spin chains \eq{a1} and \eq{A4b} do not follow
the Poissonian distribution \cite{BFGR08epl,BFGR08}.
We have already noted that, the spectrum of the presently considered
spin chain \eq{A4} 
leads to the spectra of the spin chains \eq{a1} and \eq{A4b} 
in the special cases $|m_1-m_2 | \leq 1$ and $m_2=0$ respectively. 
Hence, in the following, we shall focus on the spectral 
properties of the spin chain \eq{A4} in the case of non-zero values of 
$m_1$ and $m_2$, which satisfy the relation $|m_1-m_2 | > 1$.

For any finite values of $m_1$, $m_2$ and $N$,  
one can in principle compute the exact spectrum of the spin chain \eq{A4}
by expanding its partition function  $\mc{Z}_N^{(m_1, m_2)}(T)$ 
\eq{A25} in powers of $q$. 
Indeed, with the help of symbolic software package like Mathematica, it is 
possible to explicitly write down 
$\mc{Z}_N^{(m_1, m_2)}(T)$ as a polynomial of $q$ 
for certain ranges of $m_1$, $m_2$ and $N$. 
If the term $q^{\mc{E}_i}$ appears in such a polynomial, then 
$\mc{E}_i$ would represent an energy eigenvalue and the coefficient of 
$q^{\mc{E}_i}$ would determine the degeneracy factor corresponding 
to this energy level.
Let us denote this degeneracy factor or `level density' associated
with the energy level $\mc{E}_i$ as $\tilde{d}(\mathcal{E}_i)$.
Since the sum of these level
densities for the full spectrum is not normalized to unity, we obtain     
the corresponding normalized level density $d(\mathcal{E}_i)$ through
the relation $d(\mathcal{E}_i)= \tilde{d}(\mathcal{E}_i)/m^N$.
However, this method of computing the spectrum and 
the level density of the spin chain \eq{A4}
by using its partition function \eq{A25} is not very efficient 
for large values of $N$ (for example, using Mathematica in a personal 
computer, we can compute the level density up to 
about $N=20$ for $m_1=3$ and $m_2=1$ case). 
To overcome this problem, we consider \Eq{ANBN2} 
which gives an alternative expression of  $\mc{Z}_N^{(m_1, m_2)}(T)$  
in terms of partition functions like 
$\tilde{\mathcal{Z}}^{A,m}_{N}(T)$ associated with the scaled Hamiltonian of 
the $A_{N-1}$ type of PF chain (\ref{scal}). Furthermore, 
instead of directly using Eq. (\ref{A27b}) for expressing 
$\tilde{\mathcal{Z}}^{A,m}_{N}(T)$ in a polynomial form,  we use  
the known equivalence relation between this partition function 
and the partition function of a particular type of     
 one-dimensional inhomogeneous vertex model \cite{BBH10}.
Applying this connection with the partition 
function of a one-dimensional vertex model, which 
can be expressed as a polynomial of $q$ in a more efficient way 
with the help of Mathematica software,
we have been able to compute the spectrum and the level density  
of the spin chain \eq{A4} for comparatively large values of $N$, 
for example, up to $N=80$ with $m_1=3$ and $m_2=1$. 

In order to compare the energy level density of the spin chain \eq{A4}
with a curve like Gaussian distribution, it is needed to  
calculate the corresponding mean ($\mu$) and variance ($\sigma$). 
These parameters are related to the Hamiltonian $\mc{H}^{(m_1,m_2)}$ 
(\ref{A4}) as 
\beq 
\mu=\frac{\tr\left[\mc{H}^{(m_1,m_2)}\right]}{m^N}, ~~~~
\sigma^2= \frac{\tr \left[(\mc{H}^{(m_1,m_2)})^2\right]}{m^N}-\mu^2 \, .
\label{musig}
\eeq
Defining a parameter $t$ as  $t \equiv m_1-m_2$, and 
applying Eqs.~\eq{diag2} as well as \eq{spin4}, it is easy to find out the 
following trace relations:
\bea
 &&\tr \left[\one \right]=m^N, ~~\tr\left[ P_i^{(m_1,m_2)} \right]=m^{N-1}t,
~~\tr \left[P_{ij}\right]=\tr 
\left[\widetilde{P}_{ij}^{(m_1,m_2)}\right]=m^{N-1}, \nn \\
&&\tr\left[P_{ij}P_i^{(m_1,m_2)}\right]
=\tr \left[P_{ij}P_k^{(m_1,m_2)}\right]=m^{N-2}t, \nn \\
&&\tr\left[\widetilde{P}_{ij}^{(m_1,
m_2)} P_i^{(m_1,m_2)}\right]=\tr\left[\widetilde{P} _ {ij}^{(m_1,m_2)}
P_k^{(m_1,m_2)}\right] =m^{N-2}t, \nn \\
&&\tr\left[P_{ij}P_{jl}\right]=\tr \left[ P_{ij}P_{kl}\right]
=\tr\left[P_{ij}\widetilde{P}_{jl}^{(m_1,m_2)}\right]
=\tr\left[P_{ij}\widetilde{P}_{kl}^{(m_1,m_2)}\right]=m^{N-2},
\nn \\ 
&&\tr\left[\widetilde{P}_{ij}^{(m_1,m_2)}\widetilde{P}_{jl}^{(m_1,m_2)}\right]
=\tr\left[\widetilde{P}_{ij}^{(m_1,m_2)}\widetilde{P}_{kl}^{(m_1,m_2)}\right]
 =m^{N-2}, \nn \\
&&\tr\left[P_{ij}\widetilde{P}_{ij}^{(m_1,m_2)}\right]
=\tr\left[P_i^{(m_1,m_2)}P_j^{(m_1,m_2)}\right]=m^{N-2}t^2, \nn
\eea 
where
it is assumed that $i, j, k, l$ are all different indices.
Using \Eq{musig} along with the above mentioned trace relations, we obtain 
\beq
\mu=\left(1+\frac{1}{m}\right)\sum_{i\neq j}~(h_{ij}
+\widetilde{h}_{ij})+\frac{\beta}{2}\left(1-\frac{t}{m}\right)\sum_{i=1}^{N}\frac{1}{
y_i},  
\label{mu}
\eeq
and
\beq
\sigma^2 
=2\left(1-\frac{1}{m^2}\right)\sum_
{i\neq j}(h^2_{ij}+\widetilde{h}^2_{ij})+\frac{4}{m^2}(t^2-1)\sum_{
i\neq j}h_{ij}\widetilde{h}_{ij}+\frac{\be^2}{4}\left(1-\frac{t^2}{m^2}
\right)\sum_{i=1}^{N}\frac {1}{y_i^2}. 
\label{sigma}
\eeq
Using the identities in \Eq{iden} and also similar identities
given by \cite{Ah78,AM83,BFGR08} 
\bea 
&&\sum_{i=1}^{N}\frac {1}{y_i^2}=\frac{N(N+\beta)}{\beta^2(1+\beta)},
~~~~\sum_{i\neq j}~h_{ij}\widetilde{h}_{ij}=\frac{N(N-1)}{16(1+\beta)}, ~~
  \nn \\
&&\sum_{i\neq
j}~(h_{ij}^2+\widetilde{h}_{ij}^2)=
\frac{N(N-1)}{72(1+\beta)}\left[2\beta(2N+5)+4N+1\right] , 
\nn
\eea
we finally express $\mu$ \eq{mu} and $\sigma^2$ \eq{sigma}
 in closed forms like 
\beq
\mu=\frac{1}{2} \left(1+\frac{1}{m}\right)N^2-\frac{1}{2m}(1+t)N,
 \label{B1}
\eeq
\beq
\sigma^2=\frac{1}{36}\left(1-\frac{1}{m^2}\right)N(4N^2+6N-1)+\frac{1}{4m^2}
(1-t^2)N.
 \label{B2}
\eeq

Taking different sets of non-zero values of $m_1$, $m_2$ satisfying  
the relation $|m_1-m_2 | > 1$, and moderately large values of $N$ ($N\geq 15$), 
we find that the normalized level density of the spin chain \eq{A4}
is in excellent agreement with the Gaussian distribution (normalized to unity)
given by 
\beq
g(\mathcal{E})=\frac{1}{\sqrt{2 \pi}\sigma}
e^{-\frac{(\mathcal{E}-\mu)^2}{2\sigma^2}}. 
\label{A41}
\eeq
As an example, in Fig.~1 we compare the normalized level density
  with the Gaussian distribution  
for the case $m_1=3, ~m_2=1$ and $N=20$.
\begin{figure}[htb]
\begin{center}
\resizebox{100mm}{!}{\includegraphics{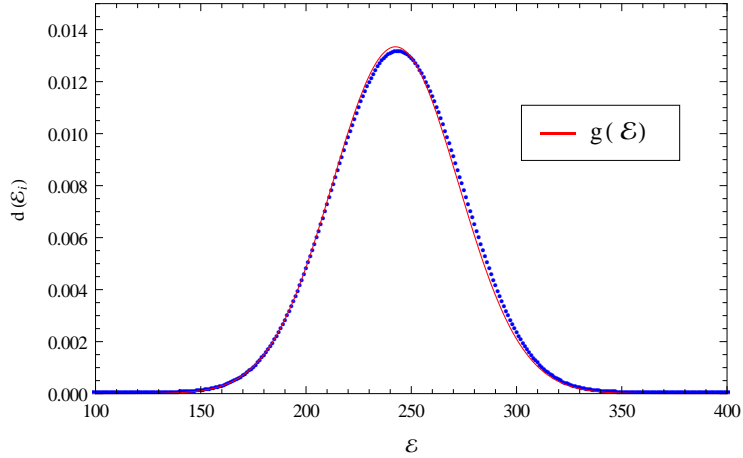}}
{\small{\caption{Continuous curve represents the Gaussian distribution 
and circles represent level density distribution for $N=20$ and $m_1=3, ~m_2=1$.
}}}
\label{test}\end{center}
\end{figure}
We also compute the mean square error (MSE) between the normalized 
level density and the Gaussian distribution for the above mentioned case
and find it to be as low as $1.73 \times10^{-8}$. 
Moreover, it is found that this MSE decreases steadily with
increasing number of lattice sites.
In Table~1 we present the values of MSE calculated    
by taking different sets of values of $m_1$ and $m_2$ for a wide range of $N$. 
\begin{table}[htb]
\footnotesize
\hspace{.9cm}
\renewcommand{\arraystretch}{1.9}
\begin{tabular}{|c|c|c|c|c|c|c|}
 \hline 
\multicolumn{2}{|c|}{\footnotesize{Sets of
Parameters}}&$N=20$&$N=30$&$N=40$&$N=50$&$N=60$\\
\cline{1-2} \hspace{.3cm}
$m_1$ \hspace{.2cm} & $m_2$&&&&&\\
\hline
$3$&$1$&$1.73 \times 10^{-8}$&$1.42 \times 10^{-9}$&$2.45 \times 10^{-10}$&$6.32
\times 10^{-11}$& $2.10 \times 10^{-11}$\\
 \hline
$4$&$1$&$1.64 \times 10^{-8}$&$1.34 \times 10^{-9}$&$2.31 \times 10^{-10}$&$5.94
\times 10^{-11}$& $1.97 \times 10^{-11}$\\
 \hline
$4$&$2$&$1.58 \times 10^{-8}$&$1.30 \times 10^{-9}$&$2.22 \times
10^{-10}$&$5.72 \times 10^{-11}$&$1.90 \times 10^{-11}$\\
 \hline
$5$&$1$&$1.60 \times 10^{-8}$&$1.30 \times 10^{-9}$&$2.23 \times
10^{-10}$&$5.74 \times 10^{-11}$&$1.90 \times 10^{-11}$\\
 \hline
\end{tabular}
\caption{ MSE for  level density  
of $BC_N$ type PF chain with PSRO (\ref{A4}).}
\end{table}

Next, our aim is to study the distribution of spacing between
consecutive energy levels for the case of  $BC_N$ type of PF spin chain 
with PSRO \eq{A4}. To this end, let us define cumulative level spacing
distribution as
\beq
P(s) = \int_0^s p(x) dx \, ,
\label{cumu} 
\eeq
where $p(x)$ denotes the probability density of the normalized spacing
$x$ between consecutive unfolded energy levels.
In order to eliminate the effect of 
level density variation in the calculation of $p(x)$,
an unfolding mapping is usually applied to the `raw' spectrum \cite{Ha01}.
For the purpose of defining such unfolding mapping, at first 
the cumulative energy level density is decomposed as the sum of a
fluctuating part and a continuous part (denoted by $\eta(\mc{E})$). 
We have already seen that,
the energy level density of the spin chain \eq{A4} follows the 
Gaussian distribution with very good approximation. Hence, for this case, 
$\eta(\mc{E})$ can be expressed through the error function as
\beq
\eta(\mathcal{E})=\frac{1}{\sqrt{2\pi\si^2}} \int_{-\infty}^{\mathcal{E}}
e^{-\frac{(x-\mu)^2}{2\si^2}} dx =
\frac{1}{2}\left[1+\rm{erf}\left(\frac{\mathcal{E}-\mu}
{\surd{2}\sigma}\right)\right]. 
\label{A43}
\eeq
This continuous part of the
cumulative level density is used to transform each
energy  level $\mathcal{E}_i,~~ i = 1 , . . . , n$, into an
unfolded energy level given by $\eta_i \equiv \eta(\mathcal{E}_i)$. 
Finally, the function $p(s_i)$ is defined as the probability density 
of normalized spacing $s_i$ given by $s_i = (\eta_{i+1} -\eta_i) / \Delta$, 
where $\Delta = (\eta_n-\eta_1) /(͑n-1)$ denotes the mean spacing
of the unfolded energy levels. 

According to a well-known
conjecture by Berry and Tabor, for the case of a quantum integrable system,
the density $p(s)$ of normalized spacing should obey the
Poisson's law: $p(s) = e^{-s}$ \cite{BT77}. However, it has been found 
earlier that a large class of quantum integrable HS and PF like spin
chains violate this conjecture and lead to 
non-Poissonian distribution of $p(s)$ 
\cite{FG05,EFGR05,BFGR08,BFGR08epl,BB09}. Moreover, the 
cumulative level spacing distributions of such spin chains
obey a certain type of `square root of a logarithm' law,
which can be derived analytically by making a few assumptions
about the corresponding spectra.  
More precisely, if the  discrete spectrum of a quantum system 
satisfies the following four conditions: \\
i) the energy levels are equispaced, i.e.,
$\mathcal{E}_{i+1}-\mathcal{E}_i=\delta$, for $i=1, 2, \ldots, n-1$,  \\
ii) the level density is approximately Gaussian, \\
iii) $\mathcal{E}_{max}-\mu, ~\mu-\mathcal{E}_{min} \gg \sigma$, \\
iv) $|\mathcal{E}_{max}+\mathcal{E}_{min}-2\mu| \ll
\mathcal{E}_{max}-\mathcal{E}_{min}$, \\
then the cumulative level spacing
distribution is approximately given by an analytic expression of the form
\cite{BFGR08} 
\beq
\tilde{P}(s)\simeq 1-\frac{2}{\sqrt{\pi}s_{max}}\sqrt{\ln \left(\frac{s_{max}} 
{s}\right)} \, ,
\label{B3}
\eeq
where 
\beq
s_{max}=\frac{\mathcal{E}_{max}-\mathcal{E}_{min}}{\sqrt{2\pi}~\sigma} \, .
\label{smax}
\eeq
We have already seen that the conditions i) and ii) are obeyed for the 
spectrum of the spin chain \eq{A4}. 
Due to Eqs.~\eq{energy2}, \eq{energy4} and \eq{nzq}, it follows that
$\mathcal{E}_{min}=N^2/m + O(N)$ and $\mathcal{E}_{max}=N^2$. Moreover,
using Eqs.~(\ref{B1}) and (\ref{B2}), one obtains the leading order 
contributions to mean and variance as
\beq
\mu=\frac{1}{2} \left(1+\frac{1}{m}\right)N^2 + O(N), ~~~~
\sigma^2=\frac{1}{9}\left(1-\frac{1}{m^2}\right)N^3+O(N^2).  \nn
\eeq
Using these leading order contributions to $\mathcal{E}_{min}$, 
$\mathcal{E}_{max}$, $\mu$ and $\sigma^2$, it is easy to check that 
the conditions iii) and iv) are also obeyed for the 
spectrum of the spin chain \eq{A4}.
Hence, it is expected that $P(s)$ would follow the analytical expression 
$\tilde{P}(s)$ \eq{B3} in the case of spin chain \eq{A4}. By using Mathematica, 
we calculate  $P(s)$ for  
different values of $m_1$, $m_2$ and for  moderately large values of $N$, 
and find that  $P(s)$ matches with $\tilde{P}(s)$ extremely well
in all of these cases. As an example, in Fig.~2 we compare $P(s)$
with $\tilde{P}(s)$ for the case of $m_1=3, ~m_2=1$ and $N=20$.
Thus we may conclude that, similar to the case
of many other quantum integrable spin chains with long-range interaction, 
the cumulative distribution of
spacing between consecutive energy levels of the spin chain \eq{A4}
follows the `square root of a logarithm' law \eq{B3} with remarkable
accuracy.
\begin{figure}[htb]
\begin{center}
\resizebox{100mm}{!}{\includegraphics{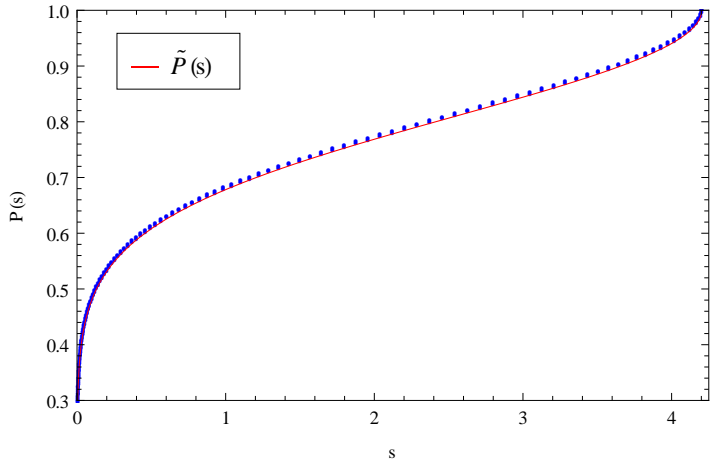}}
{\small{\caption{Circles represent cumulative spacing distribution $P(s)$,
 while continuous line is its analytic approximation $\tilde{P}(s)$  drawn
for $N=20$ and $m_1=3, ~m_2=1$.
}}}
\label{test}
\end{center}
\end{figure}

\bigskip

\noi \section{Conclusions}
\renewcommand{\theequation}{7.{\arabic{equation}}}
\setcounter{equation}{0}
\medskip
In this paper we construct the PSRO \eq{diag2} which, along with the spin 
exchange operator \eq{spin4}, yields a class of representations for the $BC_N$ 
type of Weyl algebra in the internal space associated with $N$ number 
of particles or lattice sites. This PSRO allows us to find out  
novel exactly solvable variants \eq{A2} of the $BC_N$ type of 
spin Calogero model. Taking the strong coupling limit of these 
spin Calogero models and also 
using the freezing trick, we obtain the $BC_N$ type of PF spin chains with PSRO
\eq{A4}. In one limit, these spin chains reproduce the $BC_N$ type of 
PF models studied by Enciso et. al.~\cite{BFGR08}. 
In another limit, these spin chains 
yield new SA type generalization \eq{a2} of the PF spin chain. 

Subsequently, we construct some (nonorthonormal) basis vectors 
for the Hilbert spaces of the $BC_N$ type of spin Calogero models with PSRO 
by using the projector \eq{pro22} and derive the
exact spectra of the these models by taking advantage of the fact that 
their Hamiltonians can be represented in triangular form  
while acting on the above mentioned basis vectors. Then  
we apply the freezing trick to compute the partition functions \eq{A25} 
for the  $BC_N$ type of PF spin chains with PSRO.
Furthermore, we derive a remarkable relation \eq{ANBN2} between the partition 
function of the $BC_N$ type of PF spin chain with PSRO 
and that of the $A_{N-1}$ type of PF spin chain. This relation
turns out to be very efficient in studying spectral properties like 
level density and distribution of spacings for consecutive
levels in the case of $BC_N$ type of PF spin chains with PSRO. 
We find that, similar to the case
of many other quantum integrable spin chains with long-range interaction,
the level density of these spin chains follows the Gaussian distribution 
and the cumulative distribution of spacing for consecutive levels 
follows a `square root of a logarithm' law.

Taking a particular limit of the relation \eq{ANBN2}
we obtain \Eq{A35}, which implies that 
the spectrum of the SA type generalization of the PF spin chain \eq{a2}
with arbitrary value of the parameter $\beta$ would coincide with that of the   
scaled Hamiltonian \eq{scal} for $A_{N-1}$ type of PF spin chain.  
This result is rather surprising, since the forms of the two above
mentioned Hamiltonians apparently differ from each other. 
For the purpose of making some connection between these two apparently
different types of Hamiltonians, we conjecture the 
asymptotic relation \eq{A38} between the (ordered) zero points of the 
Hermite polynomial and the generalized Laguerre polynomial. If this 
conjecture is correct, then the scaled Hamiltonian  \eq{scal} of the 
$A_{N-1}$ type of PF spin chain can be seen as a particular limit 
of the Hamiltonian \eq{a2}
corresponding to the SA type generalization of the PF spin chain.
However, we have only verified the conjecture \eq{A38} analytically 
for the case of $N=2$ and numerically for $N=3$ and $N=4$.
Therefore, finding out an analytical proof of the conjecture \eq{A38}
might be an interesting problem to study 
from the viewpoint of orthogonal polynomials.  

Finally it should be noted that, apart from the context of the    
$BC_N$ type of spin Calogero model and PF spin chain,
the $BC_N$ type of Weyl algebra plays a very important role 
in context of the $BC_N$ type of spin Sutherland model, related HS spin chain
and also for the cases of supersymmetric generalizations of these models
\cite{EFGR05,BFGR09}. 
Therefore, one can use the PSRO \eq{diag2} to construct novel exactly solvable 
variants of the $BC_N$ type of spin Sutherland model and HS spin chain 
\cite{BGF14}. Furthermore, it is also possible to construct supersymmetric 
generalization of the PSRO \eq{diag2} and apply such operator 
to find out exactly solvable variants of the supersymmetric 
spin Calogero model and PF spin chain
associated with the $BC_N$ root system \cite{BBB14}.

\vskip .85 cm 

\noi {\bf Acknowledgements}
\medskip

We thank Artemio Gonz\'alez-L\'opez and Federico Finkel 
for fruitful discussions.

\newpage 

\bibliographystyle{model1a-num-names}
\bibliography{cmprefs}

\end{document}